\documentclass[transmag]{IEEEtran}

\usepackage{latexsym}
\usepackage{graphicx}
\usepackage{amsfonts,amssymb,amsmath}
\usepackage{hyperref}

\usepackage{algorithmic}
\usepackage{textcomp}
\usepackage{xcolor}
\usepackage{import}
\usepackage{tabularx}
\usepackage{array}
\usepackage{tabu}
\usepackage{multirow}
\usepackage{siunitx}
\usepackage{booktabs}
\usepackage{times}
\usepackage{fancyhdr,graphicx,amsmath,amssymb}
\usepackage[ruled,vlined,linesnumbered]{algorithm2e}
\usepackage{mathtools}
\include{pythonlisting}
\usepackage{subcaption}
\usepackage{silence}
\usepackage{bbm}

\newcommand\Topspace{\rule{0pt}{4ex}}     % Top strut
\newcommand\Bottomspace{\rule[-2ex]{0pt}{0pt}}  % Bottom

\def\BibTeX{{\rm B\kern-.05em{\sc i\kern-.025em b}\kern-.08em T\kern-.1667em\lower.7ex\hbox{E}\kern-.125emX}}

\makeatletter
\renewcommand*\env@matrix[1][*\c@MaxMatrixCols c]{%
  \hskip -\arraycolsep
  \let\@ifnextchar\new@ifnextchar
  \array{#1}}
\makeatother

\newcommand\coolover[2]{\mathrlap{\smash{\overbrace{\phantom{%
    \begin{matrix} #2 \end{matrix}}}^{\mbox{$#1$}}}}#2}

\begin{document}
\bstctlcite{IEEEexample:BSTcontrol}

\title{UAV-Assisted Communication in Remote Disaster Areas using Imitation Learning}

\author{Alireza Shamsoshoara, Fatemeh Afghah, Erik Blasch, Jonathan Ashdown, Mehdi Bennis
% \IEEEmembership{Member, IEEE}
\thanks{This material is based upon work supported by the Air Force Office of Scientific Research under award number FA9550-20-1-0090 and the National Science Foundation under Grant Numbers CNS-2034218, CNS-2039026, and ECCS-2030047. Distribution A: Approved for Public Release, distribution unlimited. Case Number AFRL-2021-1039 on March 30, 2021.}
\thanks{A. Shamsoshoara and F. Afghah are with the School of Informatics, Computing and Cyber Systems at Northern Arizona University, Flagstaff, AZ, 86011 USA (e-mail: \{alireza\_shamsoshoara, fatemeh.afghah\}@nau.edu).}
\thanks{E. Blasch and J. Ashdown are with Air Force Research Laboratory, Rome, NY 13441, USA, (e-mail: \{erik.blasch, jonathan.ashdown\}@us.af.mil).}
\thanks{M. Bennis is with 
Centre for Wireless Communications, University of Oulu, Finland (e-mail: mehdi.bennis@oulu.fi).}}

\IEEEtitleabstractindextext{
\begin{abstract}
The damage to cellular towers during natural and man-made disasters can disturb the communication services for cellular users. One solution to the problem is using unmanned aerial vehicles to augment the desired communication network. The paper demonstrates the design of a UAV-Assisted Imitation Learning (UnVAIL) communication system that relays the cellular users’ information to a neighbor base station. Since the user equipment (UEs) are equipped with buffers with limited capacity to hold packets, UnVAIL alternates between different UEs to reduce the chance of buffer overflow, positions itself optimally close to the selected UE to reduce service time, and uncovers a network pathway by acting as a relay node. UnVAIL utilizes Imitation Learning (IL) as a data-driven behavioral cloning approach to accomplish an optimal scheduling solution. Results demonstrate that UnVAIL performs similar to a human expert knowledge-based planning in communication timeliness, position accuracy, and energy consumption with an accuracy of 97.52\% when evaluated on a developed simulator to train the UAV.

\end{abstract}

}

\maketitle

% ******************************************
% Introduction
\IEEEraisesectionheading{\section{Introduction}
\label{sec:Introduction}}

\IEEEPARstart{D}{evastating} natural disasters such as \textit{climatological} (wildfires, drought), \textit{biological} (animal plague, disease) \textit{geophysical} (volcano, earthquake), and \textit{hydrological} (avalanche, floods)  put human lives in danger. As a result, first and zero responders aim to help the people in the affected area in a timely manner by locating the survivors, repairing the damaged infrastructure, and providing communication food, medicine, etc.  \cite{erdelj2017help, rabta2018drone}. Unmanned Aerial Vehicle (UAV) networks can offer various services during or after disasters such as agile aerial assessment of impacted areas, search and rescue in harsh and hard-to-access regions, delivering emergency supplies, and acting as aerial base stations when the communication infrastructure is damaged \cite{andreeva2020supporting, pandey2020adaptive, saraereh2020performance,Afghah_INFOCOM,Islam, shamsoshoara2020aerial, afghah2020cooperative}.

UAV systems have received a lot of attention in commercial, military, government operations of telecommunication, search and rescue (SAR), surveillance, and public safety in the recent era because of their unique features such as fast deployment, wide aerial to ground point of view \cite{blasch2012wide}, and 3-dimensional mobility~\cite{sklivanitis2018airborne,chen2019artificial,shiri2019massive}. 
Several challenges such as scalability, robustness, and performance of agile response, high throughput and low latency communication entice researchers to use Unmanned Aerial Systems (UASs) in disaster relief operations \cite{mozaffari2016unmanned, mozaffari2019tutorial,Huang, shamsoshoara2019distributed, keshavarz2020real, shamsoshoara2020survey}. 

\begin{figure}[b]
\centering
\includegraphics[width=1\linewidth]{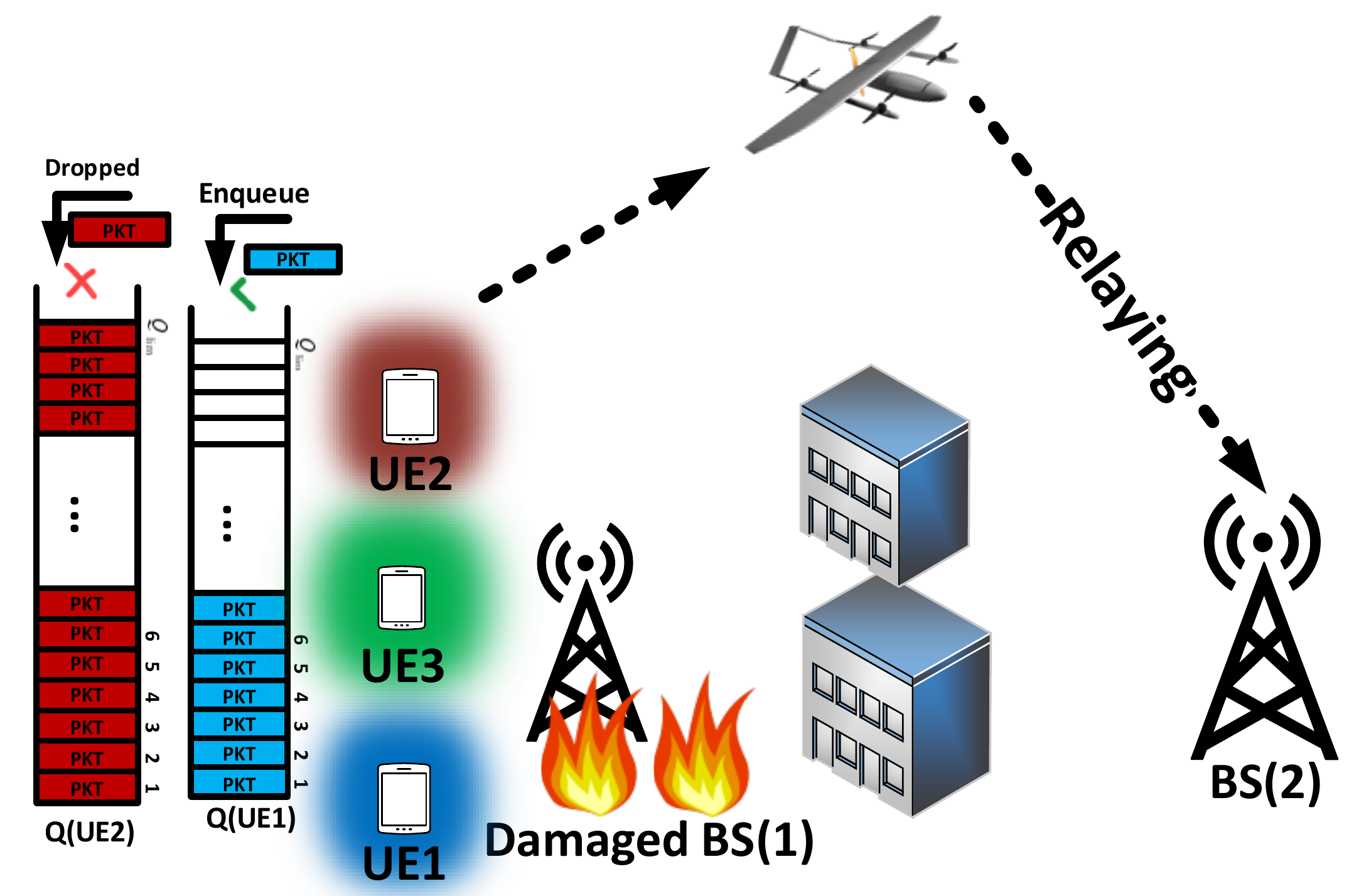}
\caption{Simplified version of the system model.}
\label{fig:system_model_simple}
\end{figure}

In this paper, we consider a situation where a natural or man-made disaster (e.g., wildfire, flood) occurs in a sparsely populated rural area with a few number of terrestrial user equipment (UEs) and completely damages the cellular base station (BS) of the Long-Term Evolution (LTE)  system to the point that it can no longer service the UEs in its region.
% , and the UEs have limited buffers to store their packets.
One solution considers a single UAV as an aerial relay in the remote disaster region to deliver the UEs' packets to an available neighbor BS before the UEs' buffers get full and the packets are dropped.
The unique features offered by the UAVs including the 3D mobility, and high probability of line of sight (LoS) transmission make them an attractive candidate for such relaying service.
This paper proposes a scheduling method for packet relaying utilizing behavioral learning where the UAV selects a UE to provide connectivity with the objectives to minimize the UE packet drop rate, decrease its own energy consumption, and increase the UEs' communication session's time.
Figure~\ref{fig:system_model_simple} shows an example situation where one BS is damaged and a UAV relays the UEs information to a neighbor operational BS. The UAV's task is to choose a UE at the right time to avoid packet dropping in queues.

%************************************Figure
%************************************Figure
% \green{We consider a vertical take-off and landing (VTOL) fixed wing UAV for this study due to its capability to perform both vertical and horizontal (hovering) movements with the efficiency of battery lifetime \cite{thamm2015songbird, tian2018hybrid, ALTIThew93:online, VectorTh84:online}.} \magenta{move this to system model}

Several recent works have studied UAV-assisted transmission scheduling for cellular users in disaster areas.
In \cite{sikeridis2018self}, a UAV-assisted non-orthogonal multiple access communication  is proposed for public safety networks, in which not all the UEs are capable of communicating with the UAV because of energy constraints or channel conditions.
% \red{where a group of UEs are clustered and located in a disaster area to collect information.} 
Hence, a minority game approach is proposed to cluster the UEs to multiple groups in a distributed manner. Next the UEs utilize a Reinforcement Learning (RL) approach to select a cluster head to join. The UAV's position is determined based on the cluster heads' location. The UEs' optimal transmission power is calculated based on a non-cooperative game theoretic model to maximize the UEs' Quality of Service (QoS).

Moreover, the studies such as \cite{shamsoshoara2019solution} and \cite{shamsoshoara2020autonomous} used Team Q-learning and a hybrid RL approach to address the task allocation and the movement of UAVs in disaster relief operations. However, in such  Markov Decision Process (MDP) problems, when the agent (i.e., the UAV) faces new system conditions that did not exist in the training scenarios, it has to start experiencing from a limited priori knowledge basis with no previous observation to interact with the environment in order to gain some knowledge. 
% \red{In real-world scenarios, it is not safe for both the UAVs and the affected people during the disaster to start the problem with no previous experience. } \magenta{does not concern this scheduling problem}
The ignorance may require spending a significant amount of time to handle the state-action space which is not appropriate for disaster relief operations. Also, in some applications such as disaster relief operations, the interaction between the agents and environment can be costly and unsafe. In addition, in real-world scenarios, it is non-trivial to define a meaningful reward function for the MDP objective functions to fully address the relation between actions and optimal policies.
In scheduling problems which involve a \textit{large} state-space and an intensive process to define a meaningful reward function for the optimal goal, an imitation learning (IL)-based approach is developed where the optimal policy in different conditions and states is determined based on real-world or simulation demonstrations from an expert that offers a more time-efficient, practical and reliable solution. In some rare cases where the expert has not experienced the state before, the agent (UAV) may face some deviation from the true path which is investigated in the last section of the simulation as well. While not applicable to our scenario, in sensitive applications, where the deviation may result in system failure, other approaches such as Inverse Reinforcement Learning (IRL) are utilized to reconstruct the expert's intention which is out of the scope of this paper.
% \red{if a human expert can demonstrate a demo of his or her optimal behavior in different conditions and states of the problem.}
Hence, the advantage of the proposed IL approach is that the UAV does not have to experience all states, but instead only mimics the expert's behavior and leverage the results to have the best outcome in each state.
In summary, the model-based optimizations are often not optimal or may lead to sub-optimal solutions. Moreover, these optimizations can be very complex. On the other hand, RL based approaches require large convergence time and heavy on-board computation. However, the proposed IL solution can offer an agile response based on the agent's observation of the expert behavior. We should note that the proposed model is defined in a way that the expert's strategy is easily reproducible for the agent by looking at the UEs' queue length as the user selection criterion, rather than finding the most optimal UE by the expert considering various factors such as physical layer  characteristics, queue length and information priority altogether since such information may not be easily accessible to the UAV agent. Therefore, as the expert's strategy in the UE selection is not the most optimal one, the agent is also not expected to find the most optimal UE but it is only rather expected to follow the expert's action without heavy onboard computations. Hence, the proposed model is not assessed by how optimal the solution is; but, we evaluate how well the proposed method can mimic the expert's action.
% \magenta{this is correct but doesn't really help the paper! If we gonna say something like this, we should start with the justification that a model-based optimization at the UAV will not be optimal and it's also complex. Then, an RL based approach also involves large convergence time and heavy computation. Then, maybe we can say the the model has been defined in simple way (with only considering important factors) from the expert's perspective to select the urgent users (so not optimal) and the UAV only aims to follow the same action without onboard computations, so the final action may not be optimal but it's expected to be similar to the expert's decision} \brown{[Alireza: That makes sense because the BC's goal is not to investigate or find the optimal solution but find the optimal policy to follow the expert.]}

In this paper, the proposed imitation learning-based solution reduces the experience's time and finds the policy taken by the expert \cite{hussein2017imitation}. To the best of our knowledge, this study is one of the first works to address the UAV-assisted transmission scheduling using an IL technique as developed in the UnVAIL solution.
The main contributions of this paper are as follows:
\begin{itemize}
    \item Develop a UAV-assisted communication using a single UAV to extend the coverage area of cellular networks or service a small number of users in remote and low traffic regions. The UAV can place itself faster in a proper sector to service the UEs;
        
    \item Devise a behavioral cloning approach that is based on a deep neural network (DNN) to reduce the level of complexity involved at the drone in the UE selection and decrease the execution time in real-time solutions for disaster scenarios;
        
    \item Showcase the dynamic techniques to determine the optimal service time based on the length of UEs' buffers to minimize the packet drop rate and save the UAV's energy, rather than the common assumption where the UAV hovers above all the UEs for a pre-determined amount of time;
        
    \item Evaluate the solution by considering the UAV's movement as a function of UE's selection decision-making based on the imitation learning approach.
\end{itemize}

The rest of the paper is structured as follows. Section~\ref{subsec:related} presents related works regarding the UAV-assisted communication and the application of IL in other domains. Section~\ref{sec:System_Model} discusses the system model and assumptions. Section~\ref{sec:Learning} introduces the imitation learning technique using the behavioral cloning technique to mimic the expert's behavior and policy. The numerical results are illustrated in Section~\ref{sec:Simulation} over a variety of metrics. Conclusions and discussions in Section~\ref{sec:Conclusion} summarize the UnVAIL solution.

\subsection{Related works}
\label{subsec:related}
While UAV-assisted communication can offer unique features for extended communication during disaster scenarios, developing an autonomous UAS which can offer reliable performance in an uncertain disaster environment still requires pragmatic design. The majority of recently published works focus on communication optimization or energy efficiency and the problem of joint path planning and packet scheduling optimizations has been scarcely studied. LTE was optimized and designed for transferring the packet data and the core network's architecture is mainly packet-switched; hence the packet scheduler (PS) has an important role in the network. Moreover, the PSs are responsible for choosing the right user at the right time for the service which affects the physical layer parameters as well \cite{capozzi2012downlink, xu2020buffer}.
% \red{While offering a reliable and flexible UAV-assisted communication in remote disaster area will be a challenging task in UAV networks, the problem of trajectory and communication optimization has been hardly studied in this domain. }
This section provides an overview of some recent works emphasizing on communication and path planning for UAV-assisted emergency in disaster scenarios.

Wu, et al., \cite{wu2018joint} considered an emergency situation in a disaster relief area where a fleet of UAVs are tasked to enhance the communication coverage and quality of service for a group of terrestrial users. The ground users are assumed to be cellular users and the UAVs are utilized as aerial BSs to enhance the downlink transmission. The problem is divided into different challenges: 1) the ground users scheduling jointly with the UAV's path planning, and 2) a power control optimization to maximize the average minimum downlink throughput rate and minimize the interference between the drones. Next, the problem is defined as a mixed-integer non-convex optimization problem. The authors used the block coordinate descent method to solve the non-convex optimization using a centralized method for a multi-UAV scenario. Although, a complex solution was used to solve this non-convex problem, the optimization technique can still approximately find the solution. Hence, the approximation of non-complex solutions only considers the location of users for the communication. However, the length of the queues or the type of applications utilized by the terrestrial users can impact the UAV communication and paths. Another drawback is that a centralized approach can be a bottleneck in the network to handle all drones.

Duong, et al., \cite{duong2019learning} proposed an optimization method for a network of relay-assisted UAVs in a disaster area, where the UAVs serve as small-cell flying BSs. The ground users are assumed to be cognitive users including primary and secondary networks. The proposed method targets the energy efficiency and the UAV's power allocation where the problem is defined as a mixed-integer optimization. The authors used a deep learning approach to solve the non-convex problem considering the power and QoS constraints for the downlink transmission. While this method contributes to reducing the execution time of finding the solution for the mixed-integer optimization, it still involves a heavy computation load and cannot be efficiently implemented at the UAV.

Game theoretical methods provide a valuable direction for UAVs' data analysis, and communication analytics which have been developed for two decades \cite{shen2008game}. For example, Koulali, et al., \cite{koulali2016green} recently proposed a fully distributed non-cooperative approach using game theory to deal with the activity scheduling for a group of UAVs in disaster relief operations. The UAVs are considered as a small set of drones to provide coverage for terrestrial users. However, guaranteeing the convergence and optimality of the proposed solution is difficult for complex problems. 
Fragkos, et al., \cite{fragkos2019disaster} considered a framework for a public safety system where the infrastructure is damaged and different agencies aim to send critical information to an emergency center using an aerial relay. The authors proposed a distributed self-optimization method for the reporting task based on two common directions of metrics \cite{blasch2004fusion} directions of \textit{Information Quality} \cite{blasch2011information} and \textit{Information Value}. A cost function is defined for each agent to show the level of information exchange between the emergency center and the agent. The authors used a non-cooperative game approach to minimize the cost function and find a level of information exchange for the agents. Next, to minimize the cost function, the authors transformed the problem into a maximization case and used a reinforcement learning approach to find the optimal information exchange level between the agencies and the UAV. In another study, Lu, et al., \cite{lu2020uav} considered a scenario of a cellular network including a UAV, multiple base stations, mobiles users, and a single smart jammer. The authors assumed that the serving base station for the mobile user is under attack by the jammer. The UAV is considered as an agent with a Deep Reinforcement Learning solution running a Q-learning approach to counteract the jammer by choosing the appropriate relaying policy to forward the mobile user information and data to another base station. The UAV does not have any information about the network topology and it only considers its experience to update its Q-values. The authors reported the performance evaluation using the bit error rate (BER) and the UAV's energy consumption rate which are obtained from both the Nash equilibrium and simulations.

A framework for a UAV-assisted network in disaster and emergency situations is studied in \cite{zhao2019uav}, where three different scenarios are investigated: i) A case of a single UAV to optimize the communication for the ground users and the UAV's flight path with an active BS; ii) Using a network of UAVs with no active BS to utilize a multihop D2D concept and extending the coverage area with a novel transceiver design; and iii) A scenario where there is no available BS and a multihop UAV relaying exchanges important information to an emergency center. In the third scenario, the relays' hovering location is optimized to enhance the performance. While this paper studied various scenarios,  all studied cases are considered for static parameters, and any small changes need a new establishment for the framework.

The authors of \cite{ferranti2019hiro} proposed an approach called ``HIRO-Net" for a disaster relief scenario where a disaster happened and there is no BS around the UEs. The approach attempted to establish a mesh network using Bluetooth for short-range communication and later used drones for each mesh for their communication. A long-range communication between the drones and emergency units is defined based on Ultra High Frequency (UHF) and Very High Frequency (VHF) links to relay the UEs' information to the emergency center. The authors considered an offline and NP-hard trajectory optimization with some predefined constraints.

In \cite{gapeyenko2018flexible}, a millimeter wave (mmWave) spectrum application to enhance the scalability and the capacity of the fifth generation (5G) mobile networks is developed. This work targets dynamic link rerouting using aerial relays and UAV networks with the aim of reducing the blockage probability on the terrestrial users. The authors proposed a mathematical framework for the UAV's speed and path planning and showed in the numerical results that the UAV-assisted framework can reduce the outage probability of the mmWave for the ground users with the aid of the UAV relays. However, it is noteworthy to mention that the mmWave technology has not been universally adopted yet and it is still highly impacted with propagation path loss, hence, the UAVs need to be equipped with highly directional antennas.

This paper proposes a learning approach called UAV-assisted Imitation Learning (UnVAIL) for the UAV drone in the remote disaster area to service the affected terrestrial users and relay their information based on their buffers' length to a neighbor BS. Unlike other previous works that considered a fixed service time for all the users, a dynamic technique is devised to determine the service time (i.e., hovering time in the UE's sector) based on the buffers' length 
% \red{and with the goal of extending the session's time between the drone and UEs.} 
in order to save the UAV's energy and service high priority UEs as needed.
Also, unlike the literature that considered static UAVs or a pre-determined path defined by a control station, in this study, the UAV's movement is a function of UE's selection decision making based on the IL approach.

% ******************************************
% System Model
\section{System Model}\label{sec:System_Model}
\begin{figure*}[hbt!]
\centering
\includegraphics[width=0.76\paperwidth]{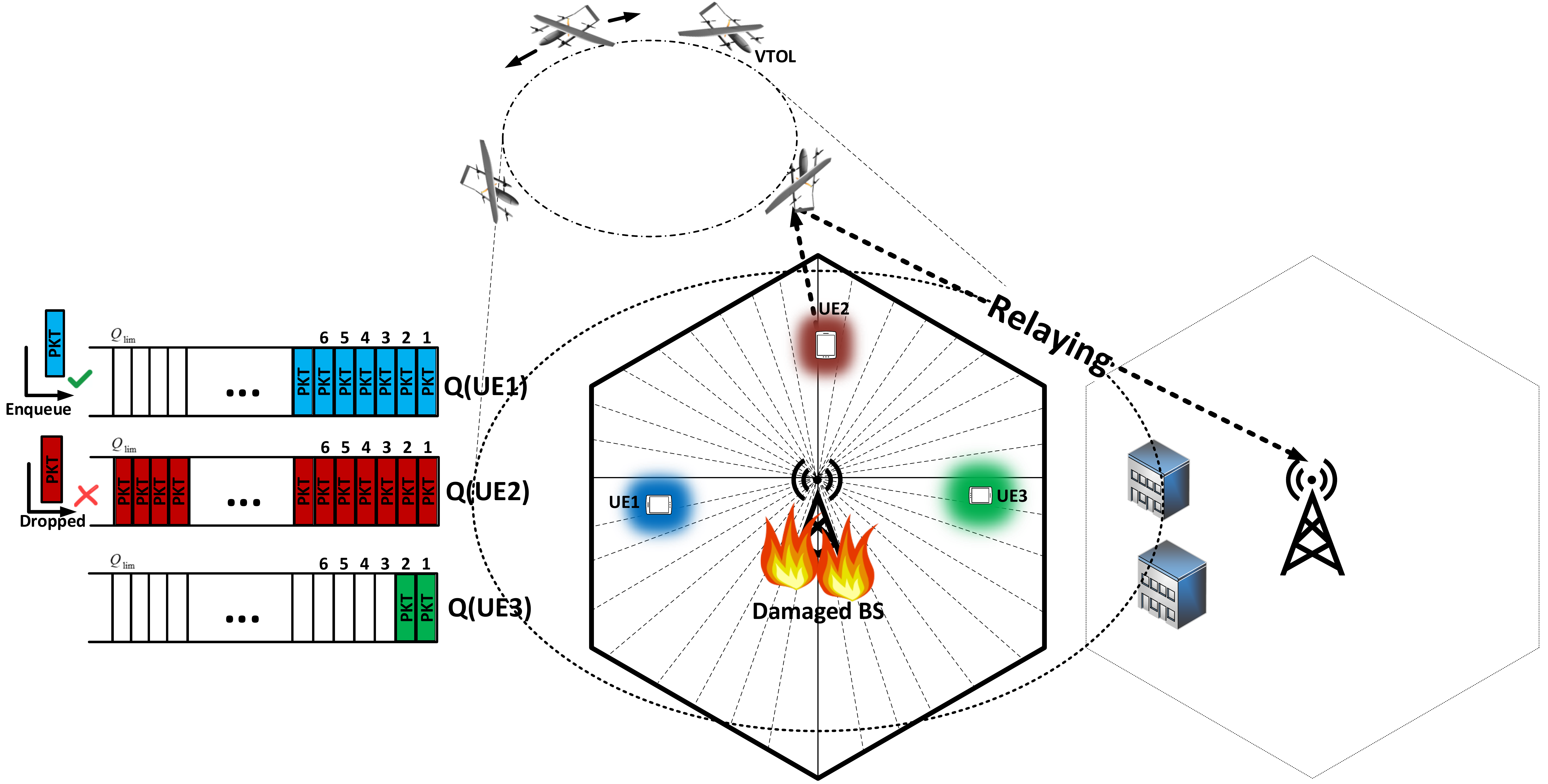}
\caption{A schematic of the proposed system model.} 
\label{fig:system_model}
\end{figure*}

Assume a network of $N$ UEs in a predefined cellular area located in a sparsely populated remote area. The UEs carry  high priority situational-awareness data such as fixed -size pictures or -duration video from the disaster to transfer. Usually, such data is not delay-sensitive and can be modeled using a constant bit rate (CBR) application for uplink transmission over a UDP agent \cite{mohankumar2013performance, rohde2013ad}. Although we assumed the CBR rate for all UEs in this system model, having a variable bit rate (VBR) does not change the approach or the methodology used in this study. It is possible to use VBR in the simulation and system definition to fit other streaming applications such as YouTube and Netflix for the bursty data traffic. The cell's base station (BS) is compromised due to the disaster's damage (e.g., wildfire). However, there are other BSs available in the neighbor regions. The BSs of the neighbor cells have access to the signal plane and information of the UEs. These BSs could provide the coverage for the UEs' uplinks. However, due to long-distance or natural phenomena such as blocking, or shadowing, the required quality of communication is not guaranteed for the data plane. Therefore, a UAV-assisted communication solution is desiring where an autonomous UAV which flies in a predefined circular path in the impacted area serves as a relaying UAV to service these UEs, as shown in Fig.~\ref{fig:system_model}. The UAV uses amplify-and-forward (AF) as the relaying technique to forward the packets to the neighbor BS. The UnVAIL solution considers vertical take-off and landing (VTOL) fixed wing UAVs due to their capability to perform both vertical and horizontal (hovering) movements, their ability to instantly change their flight direction and transition between moving and hovering, their fast speed to reach the impacted area in a short time, and their battery lifetime efficiency. These VTOLs can fly between 6 hours to 18 hours based on their battery's performance and the utilized applications. Additional options include utilizing solar UAV-charging stations if the UAVs were needed to operate for a longer time \cite{thamm2015songbird, tian2018hybrid, ALTIThew93:online, VectorTh84:online}. 

%\textcolor{magenta}{flying cell towers are usually HAPs and big UAVs that hover, you can justify that for a disaster situation you need to have UAVs with high  speed to reach the impacted area fast and provide a short-term coverage, so you are considering LAPs. You should also probably specify if you consider a quad-copter or a fixed-wing and we have conflicting assumptions here that do not fully match one of these types. Fixed-wings have longer flight-time, are more likely to keep their altitude and would be generally more appropriate to be deployed long-enough to service the UEs. We can also justify the assumption of having a circular path with fixed-wings but then we cannot have hoveing as one of the UAVs' strategies.} \brown{(Alireza: Addressed the UAV's type as a VTOL fixed wing in the introduction, I will change the figures accordingly to VTOLs, also I can mention that again in the system model.)} 
% \red{The UAV has a limited resource of battery and its task is to provide cooperative relaying between  the UEs and the BS to send the uplink data.}
All UEs can be modeled as the same CBR rate; however, this assumption can be relaxed and it does not change the methodology and the results of the simulation. The packet arrival rates for these applications are different and follow the Poisson distribution. ($\lambda _i$) denotes the packet arrival rate for UE $i$. Each UE is equipped with a queue ($Q_i$) with a limited and defined size of $Q_{lim}$ for the arrival packets. All UEs have the same queue limit. In the UnVAIL system model, the queues utilize the First-In-First-Out (FIFO) structure for the packet arrival and departure. All arriving packets are ordered in the queue based on their arrival time and if the queue reaches its maximum limitation, the arriving packets will be dropped. The system assumes that there is no age of information (e.g., time-out stamp) for the incoming packets in the system model.

%************************************Figure
%************************************Figure

Several studies in LTE-advanced relaying systems utilize the UEs' or relay's buffer level (queue length) to optimize the time, spectrum, and scheduler for the resource allocation in backhaul systems \cite{yi2012backhaul}. UnVAIL considers the queue length and queue modeling as one of the relaying priority factors for the UEs besides the packet service time, energy consumption rate, and packet drop rate \cite{wang2020energy}. The UAV acts as a server to service these queues to relay the queued information to the neighbor BS. Hence, there are multiple M/M/1 queues which the arrival time is determined by the Poisson process and the service time follows the exponential distribution with the rate of $\frac{1}{\mu}$. Each UE has a different service rate ($\frac{1}{\mu _i}$) compared to other UEs.  The amount of time that each packet spends in the queue is called the waiting time. The UAV relays the packet to the neighbor BS with respect to the service time. The moment the packet is delivered to the neighbor BS, the packet is stamped as a ``Processed'' packet.
The UAV's strategy is to avoid packet dropping at the UEs' queues. Hence, the UAV should identify a high-priority UE to forward its information to the neighbor BS. The high-priority UE is defined as a UE whose queue is getting full and the packets may start dropping; hence, the UAV should switch to this user. On the other hand, another factor for the UAV's decision-making is to save its limited energy to be able to service more UEs, where frequent switching between the UEs makes the UAV to change its location and consume more energy.

The UAV is equipped with a single directional antenna \cite{lyu2017blocking}. Hence, the UAV services a single UE at each time slot \cite{wu2018uav}. The UAV can also simultaneously perform a surveillance task in the disaster area, which often involves the UAV to fly in a circular path. Such orbit path helps avoiding the UAV's collision with other possible operational aerial drones in the area. The impacted disaster area is divided to multiple sectors, as depicted in Fig. \ref{fig:system_model}. Since a remote and sparsely populated disaster area is considered, it is assumed that the maximum number of UEs in each sector is one. 
The dimension of one sector is defined based on the UAV's antenna's beam, the UAV's altitude, and the location of neighbor cells. For the sake of simplicity, each sector is 10$^{\circ}$ of a circle with the center of the damaged BS. 
The radius of the UAV's circle is determined based on several factors including the UAV's altitude and transmission power to cover the selected UE, the location of neighbor BS, the distribution of UEs, and also considering a safe distance to avoid any collisions with other operational drones in the affected area.

The channel between the UAV and the neighbor BS is Line-Of-Sight (LoS)/Not-LoS (NLoS). The NLoS part consists of the multipath scatters from other objects between the UAV and the neighbor BS. However, the channel between the UAV and UEs is considered as the LoS one~\cite{zhang2019reflections}. %\magenta{another weak assumption for LAP. For the next paper, we need to set up the work with reasonable assumptions from the beginning. Can you revise the model to consider the common LoS/NLoS UG channel model? If not, you can at least mention the common channel model and mention LoS is selected for the sake of simplicity but it doesn't limit the applicability of the model to the general case.} \brown{[Alireza: revised it and brought a new text.]} 
It is assumed that the UAV flies at a fixed altitude. The optimal altitude is determined based on the transmission power and the trade-off between the coverage, the beam angle of the directional antenna, and the interference level to the neighbor cells \cite{khuwaja2019optimum, huang2018method}.
Despite most recent works which assume that the UAV's hover time above each user is a constant and the same for all UEs, this work defines the available time as a function of the UE's queue length. Therefore, the UAV can release itself as soon as it identifies another high-priority user rather than hovering above the UE for a pre-defined time.
The UEs operate on a single sub-band, which is assigned by the neighbor and it is used to relay the packets. 
The exchanging information between the UEs and the neighboring BSs is based on the X2 protocol in the LTE-A standard \cite{3GPP_online_subband,yaacoub2014practical}.

3rd Generation Partnership Project (3GPP) has been actively working on standardization for Long Term Evolution (LTE) to improve the efficiency of the Universal Mobile Telecommunication System (UMTS)~\cite{3GPP_online}. In the UMTS system model, the length of queues are available at the neighbor BSs. Based on~\cite{BSR_ShareTec84_online, holma2016lte, ahluwalia2011buffer, BSR_3GPP_online, BSR_3GPP_online2} from the 3GPP reports, the buffer status report (BSR) is a mechanism in which the MAC layer of the UE reports the number of packets in their buffer to the eNodeB or the BS. Several works such as \cite{kalil2014qos, rizk2016queue, rizk2015queue} used this BSR reporting method to perform the adaptive resource allocation or the quality of service (QoS)-aware scheduling based on the buffer size information. Typically, the neighbor BS shares the scheduling report with the UAV, hence the UAV has knowledge of the UEs' queue sizes. 

\begin{figure}[bth!]
\centering
\includegraphics[width=1\linewidth]{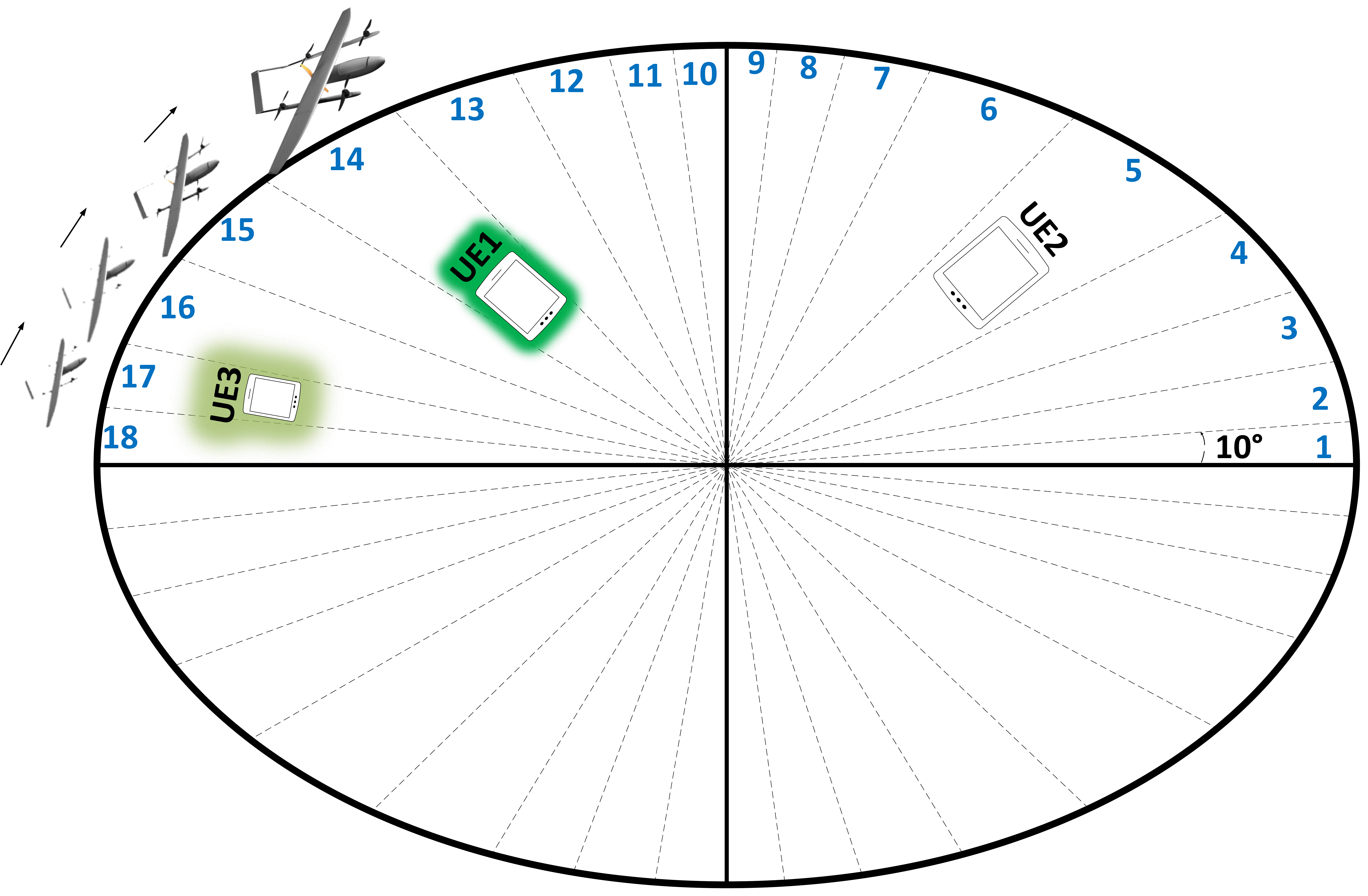}
\caption{The top view of the UAV service area, where the high priority user is changed from UE$_3$ to UE$_1$ (Sector 17 to 14). }
\label{fig:system_model_top}
\end{figure}

%\magenta{my other question is that why the UAV should service only one single user? A more realistic assumption was to define grids or sectors and assume that the UAV services all the UEs in the region to maximize the number of serviced users. We can still say that the grid/sector selection would be performed based on the high priority UE as defined in your model but would be more efficient to service all the existing users there. One way to justify your model is to say despite most the work which assume a fixed hover time to service a grid/sector, in your model, you consider an sparse rural region where you define the hover time as a function of the UE queue and the UAV is released as soon as it is done with this user instead of hovering there for a pre-defined time to service other potential users. For a sparse network, this seems to be more reasonable.} \brown{(Alireza: Agree, Added another paragraph at the beginning of the introduction to address this part).} 
Figure~\ref{fig:system_model} demonstrates the topology of the system including the relay UAV, neighbor BSs, and the UEs. The UEs are mobile in each sector. Their mobility pattern follows the Brownian motion with a constant velocity. The UEs have random directions based on the Brownian pattern but the expected location for each UE is the same allocated sector. %\red{Hence, from the beginning to the end, they are not changing their sectors.} \magenta{why do we need to assume they are fixed, we don't have any PHY later modeling and the UAV only move toward the sector not the exact location. it also does not follow the UE. How about we say the UEs have low/medium mobility and do not go out of the sector during the service time. If so, it would be good to provide some real numbers on the dimensions of the sector based on the cell dimension and sector angle and estimate UE velocity (considering some random mobility models)} \brown{Alireza: agree with you, I will add that mobility related paragraph for the UEs to explain that they are moving randomly in their sectors.} 
The UEs are located in random locations across the cell. The cell is divided into different sub-areas or sectors based on the UAV movement operation. In the scenarios of interest, the primary BS of the cell is damaged. The UAV's path is pre-defined based on a circular track. The UAV can move clockwise, counterclockwise, or remain (e.g., hover) in a fixed location. The action related to moving towards another UE consumes more energy compared to hovering in a fixed location to continue serving the same UE \cite{yang2018energy, zeng2017energy}. 
The UAV plane is also divided into different angles. For instance, in Fig.~\ref{fig:system_model_top}, the circular plane of the UAV consists of 36 sectors which each fragment is 10 degrees. In each movement action or step, the UAV flies 10$^\circ$ clockwise or counterclockwise. If both the UAV and the UE are located in the same sector, then the Euclidean distance between the two entities would be minimized. Although the service time for each packet is modeled based on the exponential distribution, the larger distance between the UAV and the UE can increase the service time as an additional delay. Figure~\ref{fig:system_model_top} shows the top view of UEs and the UAV in a sample circular plane. The UAV starts moving clockwise from sector $17$ to sector $14$ since the high-priority user is switched from $UE_3$ to $UE_1$. 

%************************************Figure
%************************************Figure

$Q_i$ denotes the queue size of the $i^{th}$ UEs. The UEs with more occupied queues are likely to experience a packet drop if the UAV cannot service them at an appropriate time. For instance, in Fig.~\ref{fig:system_model}, two packets, $PKT_1$ and $PKT_2$, arrived for $UE_1$ and $UE_2$ accordingly. The $PKT_1$ is queued since the queue for the $UE_1$ was not full. However, the $PKT_2$ is dropped since the $Q(UE_2)$ was full at that moment. 
Therefore, in the proposed model, the UAV's objective is to choose the high-priority UE to minimize the number of packet drops. Then, the UAV moves toward the selected UE's sector to minimize the service time. It is worth noting that although moving between the UEs to service the high-priority UE at any given time can reduce the rate of packet drop; however, such frequent switches come with a high energy consumption to change its location. Thus, the UAV faces a trade-off in its decision making between servicing the high priority UEs to minimize the packet drop rate and saving its energy.

In a nutshell, the problem statement based on the Energy/Delay Throughput (EDT) utility function is as follows:

% *****************************************Equation
\begin{align}\label{eq:util}
& \max_{\mathbbm{1}_J} \quad\quad EDT_f(\mathbbm{1}_J)
\\
& \textnormal{s. t. } \quad\quad\ \ EDT_f(\mathbbm{1}) = 
\frac{L * \sum\limits_{j=1}^{K} D_{f, j}(\mathbbm{1}_j)}{(\sum\limits_{j=1}^{K}t_{S_{f, j}}(\mathbbm{1}_j))} \ \times 
\\
\nonumber
&
\frac{1}{(1 + {\sum\limits_{j=1}^K\sum\limits_{i=1}^N\text{\#pkt drop}}(\mathbbm{1}_j)_{f, j,i}) *  (\sum\limits_{j=1}^{K}(e_{t_f, j} + e_{m_{f, j}}))},
\\[12pt]
&  \quad \quad \quad \quad t_S(\mathbbm{1}_j) \sim Poisson(\mu) * \alpha(\mathbbm{1}_j),
\\[12pt]
& \quad\quad\quad\quad \alpha(\mathbbm{1}_j) \propto \textnormal{Scale(Distance(UAV-UE(}\mathbbm{1}_j\textnormal{)))},
% \\
% & \quad\quad\quad\quad 
% & \textnormal{minimize } e_{sub_i}
% \\
% & \textnormal{minimize } t_{S_i}
\end{align}
% *****************************************
%\magenta{another comment you will probably get is that the position of the UAV also plays a role in the QoS for packet delivery to the BS while we only consider the distance to the UE in finding the movement strategy for the UAV. When talking about selecting the optimum altitude, you need to mention the transmission power of the UAV is selected in the way that it guarantees the packet delivery during the second hop of the packet relaying for the largest distance over the circle from the neighbor BS. For the next work, considering the power transmission of the UAV should be a decision factor. Also think about considering a multi-hop UAV network to cooperatively deliver the packets to the nearest BS, where multiple BSs may be damaged  }
where $\mathbbm{1}$ is the one-hot indication vector for the UE selection, $J$ is the set of all events in one frame from 1 to $K$ ($j \in J = \{1, 2, \dots, K\}$), $EDT_f(\mathbbm{1})$ is the EDT of the $f^{th}$ frame based on the $j^{th}$ UEs selection vector, $L$ is the number of bits or the packet length, In case that VBR is used for the variable data traffic, $L$ is not a constant anymore and it varies for each event and each UE based on their application, $L_{j, i}$. However, this does not change the study's intention to reduce the packet drop rate and the energy consumption rate. $K$ is the number of events in each frame.
$D_{f,j}(\mathbbm{1}_j)$ is the total number of delivered packets to the BS at $j^{th}$ event in frame $f$ based on the $j^{th}$ UEs selection vector. \textit{\#pkt drop}$(\mathbbm{1}_j)_{f,j,i}$ is the total number of dropped packets at the $f^{th}$ frame and $j^{th}$ event for $i^{th}$ UE based on the indicator variable. $t_{S}$, $e_{t}$, and $e_{m}$ are the packet service time, transmission energy consumption, and mobility energy consumption, respectively. $\alpha$ is a punishment factor which increases the service time because of the distance. \textit{Scale($\mathcal{R}$)} $\rightarrow$ $\mathcal{R}$ is a function which maps the distance between the UAV and the UE to a limited range of $[1,2]$. 
A simple scale function of $f = \frac{dist}{max\_dist} + 1$ is defined to map the distance to the range [1,2]. The function is defined in a way that if the UAV and the UE are in the same sector, the scaled output is 1 and if the UAV and the UE are in the longest distance from each other (e.g., sectors 1 and 19), the scaled output is 2. The scale function is defined based on the simulation and numerical results from the experiments. Although different scale functions could be used, this one is more appropriate for the emulation interaction and trade-off between the service time and switching to another UE. If the lower limit of the scale function is less than "1", then there is a contradiction with the assigned service time based on the exponential distribution. We tried different upper limits for the scale function and we observe that values more than 2 significantly affect the pack drop rate.

In summary, the UAV's action such as choosing the proper UE at a right time affects the number of dropped and delivered packets. Also, choosing the right time to switch to other UEs can save the UAV's energy. Moreover, the UAV's movement action based on the high-priority UE keeps the service time low. If the service time for a UE increases, the UAV remains busy to deliver the packets for the UEs; therefore, it is more likely for other UEs to have larger queues. The optimization variable $\mathbbm{1}_J$ in each frame is determined by the expert knowledge based on the expert's experience. Next, the UAV wants to find the proper indicator function using the UnVAIL approach. Section~\ref{sec:Learning} explains the proposed UnVAIL approach to address the distributed UAV challenge.

\section{Imitation Learning: Behavioral Cloning}\label{sec:Learning}

% \red{Introduction to Apprenticeship Learning (AL) and Imitation Learning (IL)
% After explaining the IL and AL. Jump to how we can map the state and action space to our problem. 
% What it the expert Policy (Gamer) based on our system model?} 
The operation of autonomous agents (e.g., robots or self-driving vehicles, UAVs) in uncertain environments involves complex decisions and is often time consuming \cite{blasch2019methods}.
% \red{such as UAVs need complex programming methods to operate in an uncertain environment regarding the decision making. In some cases, this task is deeply complex or time consuming.}
The objective of \textit{Imitation Learning} (IL) or \textit{Apprenticeship Learning (AL)} is to enable the agent to mimic the human experts' behavior through training scenarios obtained from real or simulation demonstrations. 
% \red{ However, human experts such as pilots know how to interact and demonstrate the proper and optimal trajectories to the UAVs based on the dynamic behaviour of the environment even they do not perform the behaviours programming to the machines or UAVs.}
Later, these training data and  optimal obtained trajectories learned from the demonstrations are used to model the expert policy for the agent's test scenario. In most disaster relief scenarios, there is a need for an agile system that can take care of high priority data. This agile system needs to be implemented on a low computational drone's computer. IL methods usually do not need complex reward function for implementation and they can be implemented with low tensor allocation on minicomputers for fast decision making using the expert's training data.

We like to note that in this study, the trajectory refers to a set of agent's states and actions- not the UAV's trajectory or movement route. The trajectory set of the state and actions is being collected and recorded in a local dataset to be trained later by the drone and the trained model can be utilized in a real-time scenario.
In this study, the state is the number of packets in queues for all UEs. For instance, if there are five UEs in the area and each UE is equipped with a limited queue with a length of 199 packets, then the possible number of states for the problem is $200^5$ = 320,000,000,000 combinations. The action will be chosen by the UAV which affects the state. The UAV chooses the UE at the right time to relay its packet to avoid the packet dropping while noting the energy involved in reaching this user. If there are five UEs in the scenario, then the UAV has five possible actions for the user selection. The policy maps the future action to the current state of the problem. In this problem, the expert and the imitation model are the possible options for the policy.
% \red{This concept is called Imitation Learning (IL) or Apprenticeship Learning (AL).}
% \red{In IL, the human or in our case the pilot demonstrates the decision making to the UAV} 
In the proposed IL-based approach, an intelligent agent, i.e. the UAV, learns optimal decisions based on imitating the expert's decisions in extensive simulation scenarios  \cite{abbeel2010autonomous,pomerleau1989alvinn,osa2018algorithmic,yu2020intelligent,lee2019learning}.

\subsection{The expert policy}
In this study, the expert is a computer with high computation capability that utilizes a human experience to optimize the objective function of the EDT in different scenarios in a limited time and generate data for the deep imitation model (UnVAIL). Then, UnVAIL tries to mimic the expert's policy and since the expert's policy is complex to recover, the expert's demonstrations are being observed for the drone's imitation model.
%\red{In this study, the expert is a human who has knowledge about the problem definition and objective function of (\ref{eq:util}) in different scenarios. The expert is a person in the operation center who has information about the drones and communication and he/she controls the drone (the emulator) for a limited time to collect data for the deep imitation model (UnVAIL).}

\begin{figure}[hbt!]
\centering
\includegraphics[width=1\linewidth]{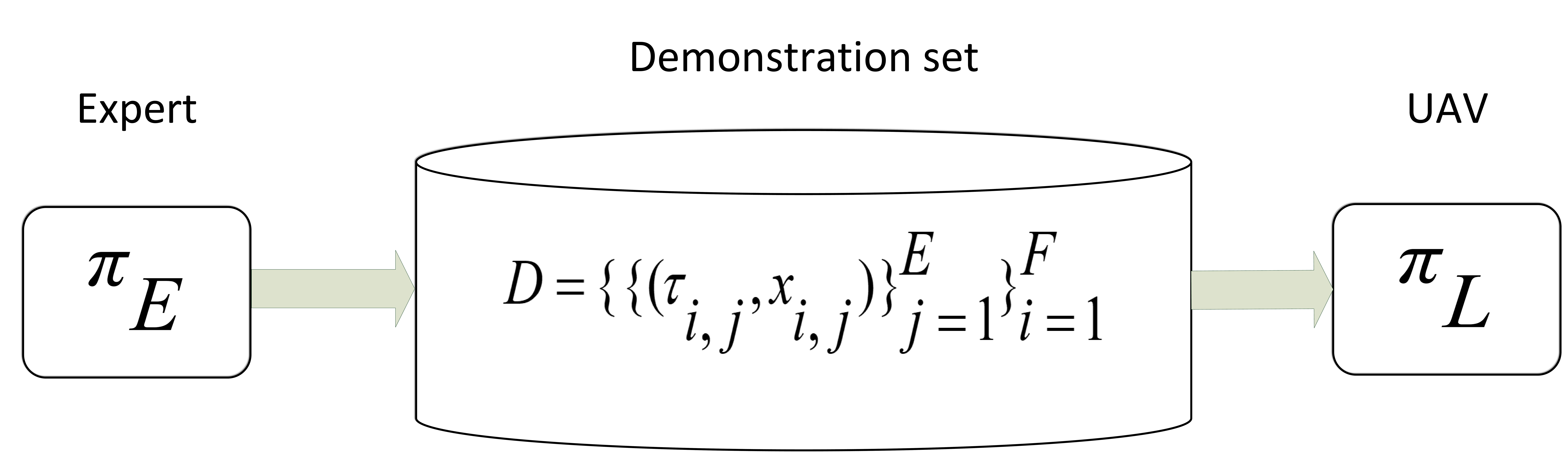}
\caption{Imitation learning diagram.}
\label{fig:imitation_learning_diagram}
\end{figure}

In general, IL methods can be categorized into three main groups: (1) behavioral cloning (BC), which is defined as directly mimicking the expert trajectories, (2) Dataset Aggregation (DAgger), and (3) apprenticeship learning via inverse reinforcement learning (IRL), in which the agent learns the hidden purpose or reward function of the expert from demonstrations. The BC's implementation does not require a  high capability system and it only requires the expert for a few trajectories at the beginning; however, any false decision compared to the expert's data may result into a big deviation. DAgger assumes that the expert is always available for those scenarios where the state has not been seen before to mitigate those deviations and improve the optimal policy, which is not feasible in many applications. In this study, the expert is not available all the time for the agent (UAV). 
% \red{IRL tries to learns the reward function by itself in order to update and improve the optimal policy in every iteration}.% \magenta{it would be more accurate to state all the three categories of IL, direct policy learning and IRL and discuss the pros and cons of each one to justify why BC is selected.} \brown{[Alireza: added a text above.]} 
In this paper, behavioral cloning approach is selected as it does not rely on the availability of the expert and involves low computation at the agent for an agile response in disaster operations. The proposed BC method replicates the expert's decisions by the UAV. Also, it is assumed that the UAV does not attempt to learn the forward model of the dynamic environment and it only tries to regenerate the demonstrated behavior by learning the policy, hence the approach in the UnVAIL solution is model-free. At the end of the demonstration, the expert dataset $\mathcal{D} = \{(\mathbf{x}_t, \mathbf{s}_t, \mathbf{a}_t) \}$ is available for the UAV. Here $\mathbf{x}_t$ is the ``context'' \cite{snidaro2016context} which stands for the initial condition or the trigger function for the next iteration. In the system model, the $\mathbf{x}_t$ is the departure of one packet from one of the UE's queues. $\mathbf{s}_t$ is the system's state at time $t$ and $\mathbf{a}_t$ is the taken action based on the observed state $\mathbf{s}_t$. The expert utilizes its policy $\pi_E$ to decide on an action: 
% *****************************************Equation
\begin{align}\label{eq:expert_policy}
\mathbf{a} = \pi_E(\mathbf{x}_t, \mathbf{s}_t), 
\end{align}
% *****************************************Equation
where, $\pi_E$ is the expert policy. The UAV uses these demonstrations to estimate the expert's policy and behavior based on Fig.~\ref{fig:imitation_learning_diagram}.
%************************************Figure
%************************************Figure
In Figure~\ref{fig:imitation_learning_diagram}, $\tau$ is the trajectories of state-action set, $E$ and $F$ are the number of events in one frame and total number of frames accordingly. $\pi_L$ is the estimated policy by the learner (UAV). It is assumed that trajectories are fully observable as a set of states and actions when the expert generates data using a simulator:
% *****************************************Equation
\begin{align}\label{eq:trajectory}
\tau = [\mathbf{s}^{(0)}, \mathbf{a}^{(0)}, \mathbf{s}^{(1)}, \mathbf{a}^{(1)}, \dots, \mathbf{s}^{({E\times F})}, \mathbf{a}^{({E\times F})}] 
\end{align}
% *****************************************Equation

After explaining the nature of IL, the action set contains two different vectors: i) $\mathcal{A}_1 = \{0, 1, ..., N-1\}$ denotes the indices for the UEs. One of the factors for the UAV to consider in choosing the next user to service with relaying is the system's current state which refers to the number of packets waiting in the queue. We like to note again that the proposed user selection process depends on multiple factors based on the defined energy-delay throughput (EDT) utility which considers the packet delivery, packet drop rate, and service time in the problem formulation. % \magenta{this sentence is confusing and may let the reviewer assume that the user selection is only based on the state information,aka the queue length, which is a simple decision making with not need for IL. However, the selection is based on the EDT} \brown{[Alireza: agree with you, added another point here about the EDT.]}
ii) $\mathcal{A}_2 = \{0, 1, 2\}$ denotes the movement actions based on the chosen UE and the current location. $0$ and $1$ notations refer to moving one sector in a clockwise or counterclockwise direction, accordingly. $2$ stands for hovering at the same location and sector without changing the angle. It is also noted that the movement action is chosen based on the selected user and it is determined based on the learning algorithm.

In the UnVAIL system model, the state space is a combination of UEs and length of the queues for all UEs, and the index of the active UE. The state feature space can be shown as a matrix with two dimensions. The row dimension stands for an occurred event in a specific frame and the column dimension shows the number of features in one state. Based on the state features, the high-priority user is chosen in the output as the first action and then the movement action is chosen based on the relative distance between the UAV's location and the high-priority user and the predefined path of the UAV. These two matrices of the state-features and taken actions are shown in (\ref{eq:sample_state}) and (\ref{eq:sample_action}) for an example scenario with three UEs.  

\vspace{5mm}
% *****************************************Matrix
% \begin{flalign}\label{eq:sample_state}
% % \resizebox{1\linewidth}{!}{$
% \phi = &
% \begin{bmatrix}[ccc|ccc|ccc|c]
%   \coolover{\textnormal{Q lengths}}{54 & 55 & 89} & \coolover{\textnormal{Distance}}{10 & 11 & 1} & \coolover{\textnormal{Direction}}{0 & 0 & 1} & 1 \\
%   55 & 57 & 88 &  9 & 12 & 0 & 0 & 0 & 2 & 2 \\ 
%   \vdots & \vdots & \vdots & \vdots & \vdots & \vdots & \vdots & \vdots & \vdots & \vdots\\ 
%     170 & 70 & 62 & 0 & 5 & 12 & 1 & 1 & 0 & 0
% \end{bmatrix},
% % $}
% \end{flalign}
% *****************************************Matrix

% *****************************************Matrix
\begin{flalign}\label{eq:sample_state}
% \resizebox{1\linewidth}{!}{$
\phi = &
\begin{bmatrix}[ccc]
   \coolover{\textnormal{Q lengths}}{54 & 55 & 89}
%   \coolover{\textnormal{Distance}}{10 & 11 & 1} &
%   \coolover{\textnormal{Direction}}{0 & 0 & 1} & 1 
   \\
   55 & 57 & 88 \\ 
   \vdots & \vdots & \vdots \\ 
    170 & 70 & 62 
\end{bmatrix},
% $}
\end{flalign}
% *****************************************Matrix

% *****************************************Matrix
\begin{flalign}\label{eq:sample_action}
(\mathbf{a}_1, \mathbf{a}_2) \in (\mathcal{A}_1, \mathcal{A}_2) = 
\begin{bmatrix}
   2 & 1 \\
   2 & 2 \\ 
   \vdots & \vdots \\ 
   0 & 2
\end{bmatrix}
\end{flalign}
% *****************************************Matrix

In (\ref{eq:sample_state}), the columns show the queue lengths of the UEs. In (\ref{eq:sample_action}), the first column shows the UE that the UAV chooses to service and the second column is the selected direction to move based on the selected UE.

% *****************************************Algorithm
\begin{algorithm}[hbtp]
\SetAlgoLined
% \KwResult{Write here the result }
 \textbf{Initialization:}\\
 Set the path for the UAV\\
 Set the initial location for all UEs and the drone\\
 Set the different arrival rates and sample rates for UEs ($\lambda_i, \frac{1}{\mu_i}$)\\
 
 \For{\textnormal{all} Runs}
 {
    Initialize all arrays, queues to zero, also set the battery to the initial value\\
    \For{\textnormal{all} Frames}
    {
        Generate the arrival and service time for all packets and all UEs in one frame ($t_A, t_S$) \\
        \For{\textnormal{all} Events}
        {
            \For{\textnormal{all} UEs}
            {
                Compare the current time with arrival time and en-queue incoming packets
            }
            Store all current queue states for the deep NN ($\phi, \tau$)\\
            Store all directions and distance information\\
            Use the expert knowledge to select the best UE ($\pi_E$)\\
            Use the GPS location to update the direction ($a_2$)\\
            Store the selected UE for the deep NN\\
            Release the packet from the chosen queue,
            service the packet, update the distance, residual battery, and the current event time\\
            % \eIf{Random $<$ $\epsilon$}
            % {
            %     Choose random actions for all UAVs\\
            %     Update the location\\
            %     Update the task
            % }
            % {
            %     Choose action based on best Q-val\\
            %     Update the location\\
            %     Update the task
            % }
            Save all expert knowledge and information in the database ($\tau$)\\
        }
    }
 }
% % %  \While{While condition}{
% % %   instructions\;
% % %   \eIf{condition}{
% % %   instructions1\;
% % %   instructions2\;
% % %   }{
% % %   instructions3\;
% % %   }
% % %  }
 \caption{Expert behavior algorithm}
 \label{algo:expert}
\end{algorithm}
% *****************************************Algorithm

Here, to access the expert knowledge, a simulator is designed for the expert to generate the training data \cite{github:code_IL_BC2020}. The simulator generates sample arrival time for the incoming packets and queues the packet in UEs' buffers. The expert uses its knowledge to observe the states of queues, locations and then decides about the active user to service and the UAV's next movement action. The expert's goal is to avoid the packet dropping or lower its rate based on the UEs' queue states. The expert's decision is based on his/her experience which is a complex model and hard to define for the UAV. All generated states and taken actions are gathered and stored for the purpose of behavioral cloning. A short video of the designed simulator including how the expert handled the problem is available on YouTube~\cite{youtube2020_imitation}. Also, the algorithm which generates the expert trajectories is presented in Algorithm~\ref{algo:expert}.

\subsection{Learner (UAV) policy}
One simple approach to find the right UE could be  sorting all state-action data in a look-up-table, then using search algorithms to find the appropriate action based on the observed state. However, there are some drawbacks with this approach such as the fact that the search algorithms are slow and also if the observed state does not exist in the table, then the returned action would be none. As a result, we choose a learning algorithm to find or estimate the proper action based on the observed state and the training data.
%************************************Figure
\begin{figure}[hbt!]
\centering
\includegraphics[width=1\columnwidth]{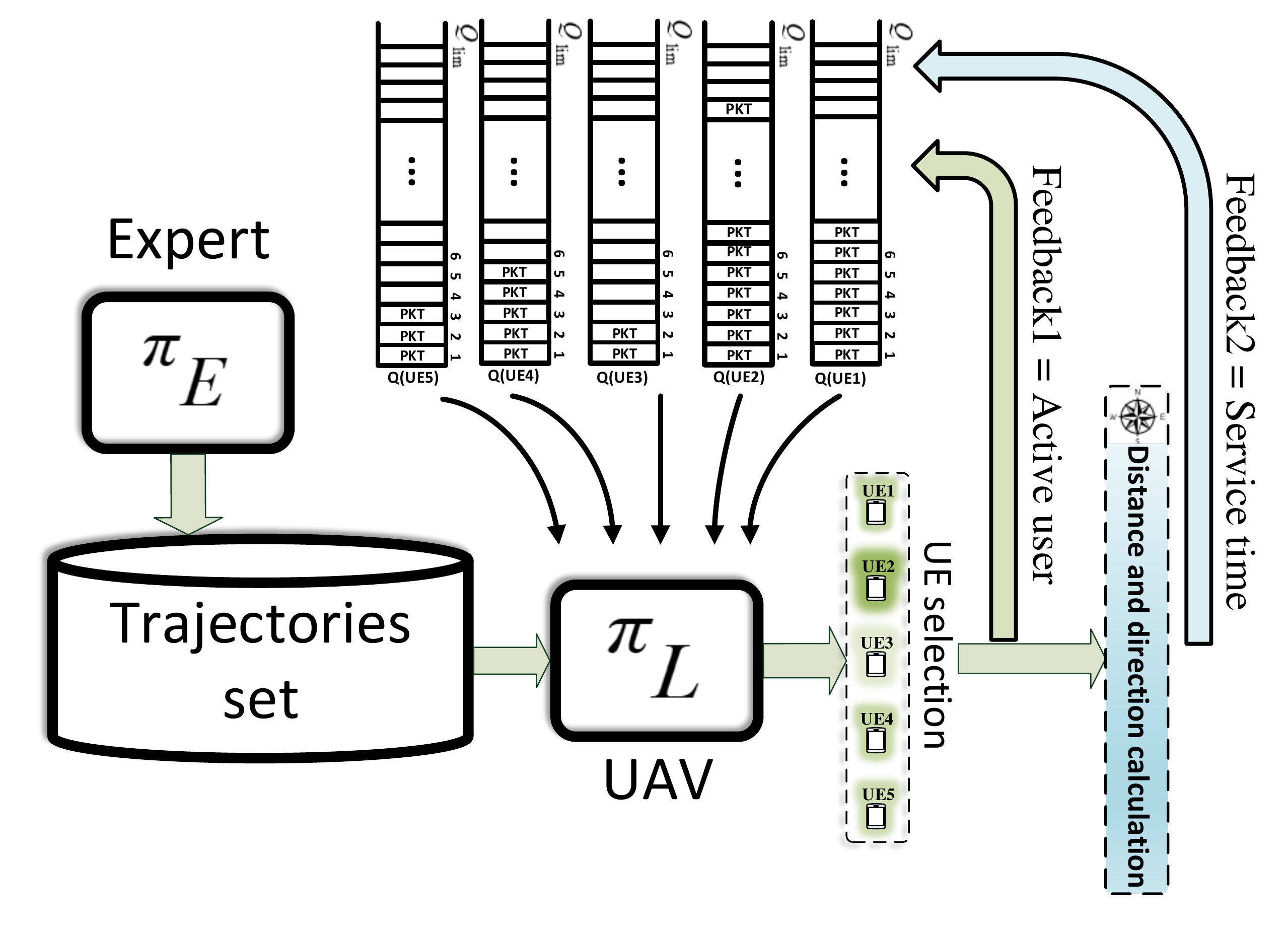}
\caption{UAV decision making model.}
\label{fig:learning_model}
\end{figure}
%************************************Figure

After generating the training data using the simulator and the expert knowledge, we have to train the data to clone the expert's behavior in the test scenario. To train the data and mimic the expert's behavior, we defined the problem as a supervised learning. More specifically, this supervised learning can be categorized as a multi-class classification problem. A learner model or behavioral model can be used to identify the right UE to service based on the queue states. We develop a deep learning model for the UAV to select the UE to deliver its packet from its queue to the base station. Then based on the selected UE, the UAV decides on the movement action based on the distance and direction information in the current state. We assumed that the GPS coordinates of the UEs are available from the base station report at each time interval. And the UAV updates its current states based on the recent changes and movements in the environment. The UAV's goal is to reduce its distance to decrease the service time. Since higher service time makes it possible for the UEs to have a longer queue. Fig.~\ref{fig:learning_model} shows the imitation learning model including the expert trajectories, the deep learning model, and the mobility selection. The user selection and service time will affect the queue lengths of the UEs as a new feedback from the previous state. $\pi_{E}$, the expert policy is used in the training phase to generate the trajectories set. All training data is used to model $\pi_{L}$, the learner policy. Hence, in the real test scenario, those trajectories data are not required. However, the current state of the queues is fed into the model, then the behavioral cloning model estimates and classifies the output by choosing a UE. We should note that in behavioral cloning models as adopted here, during the test scenario, the UAV does not have access to the expert knowledge anymore. 
% \red{Deep learning model was computationally expensive in the past. But nowadays, with the advent of GPUs and TPUs, it becomes less time consuming for the training phase. Also this study is not about the training performance; it is about the comparison between the imitated model and the expert behaviour.}

\begin{figure*}[hbt!]
\centering
\includegraphics[width=0.76\paperwidth]{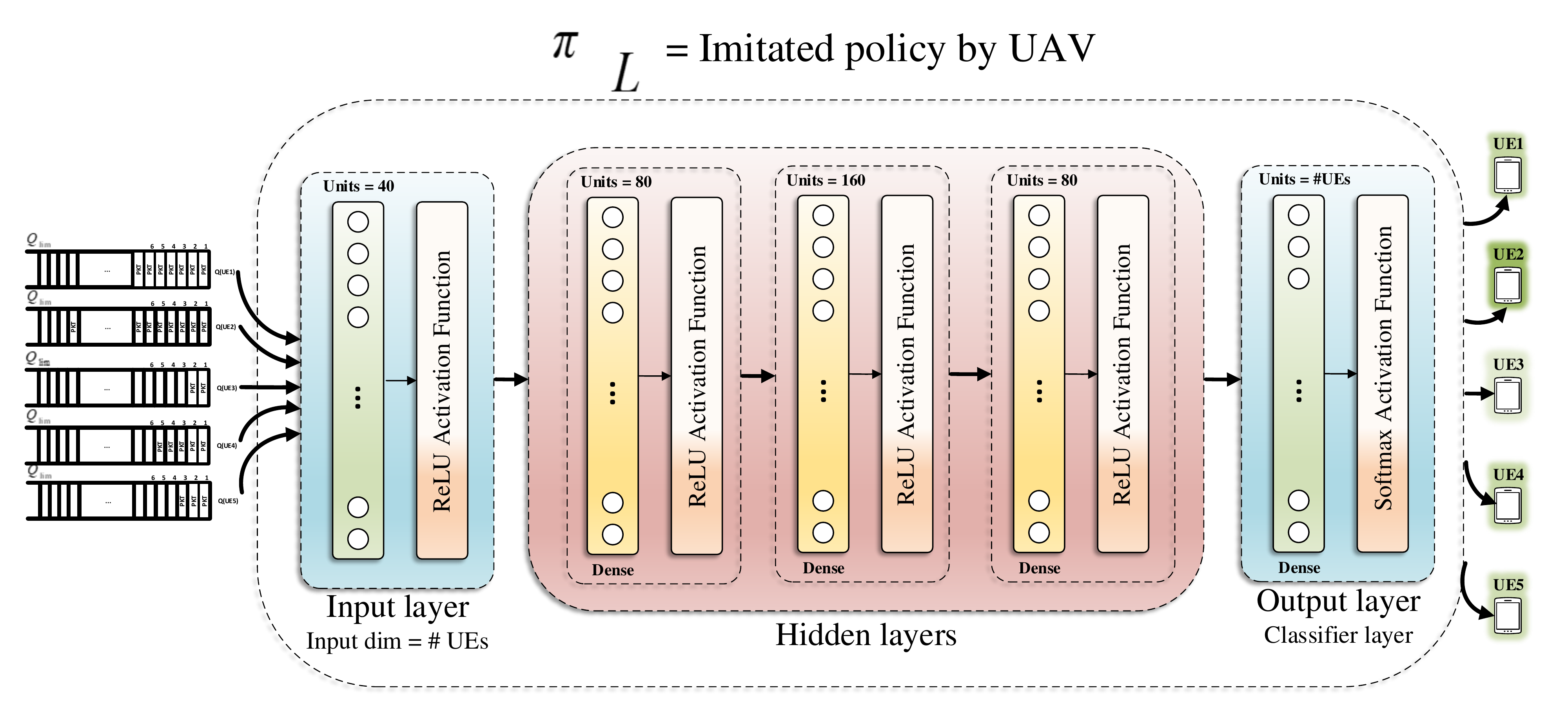}
\caption{Deep learning model for the imitated policy.}
\label{fig:neuralnetwork}
\end{figure*}

The developed deep learning model is shown in Fig.~\ref{fig:neuralnetwork}, which consists of an input layer with 40 units of neurons, where the input dimension depends on the number of UEs. The first, second, and third hidden layers have 80, 160, and 80 neurons, respectively and they are all dense layers. All input and hidden layers include a Rectified Linear Unit (ReLU) \cite{li2017convergence} layer, the output layer is a dense layer in which the number of neurons is equal to number of classes and in our scenario, it is equal to number of UEs. Since we defined this model as a classification problem, a softmax activation function~\cite{Softmaxf21:online} is used as the last component in the output layer to identify the UE in need for service. The equation for the softmax function is shown in (\ref{eq:softmax}). Algorithm~\ref{algo:training} provides more explanation about the training phase. 
% *****************************************Equation
\begin{align}\label{eq:softmax}
\sigma(UE = j | \theta(i)) = \frac{e^{\theta(i)}}{\sum\limits_{j = 0}^{K} e^{\theta_j(i)}} \qquad \textnormal{for} \ i = 1, \dots, K
\end{align}
% *****************************************Equation
In (\ref{eq:softmax}),  $K$ is the number of UEs. $\theta = (\theta_1, \dots, \theta_K) \in\mathbb{R}$ is the set of output values from the deep learning model based on the network weights and the input variables which are the queue states in this system model. These $\theta$ values are mapped to the predicted UE based on the softmax function.

%************************************Figure
%************************************Figure

% *****************************************Algorithm
\begin{algorithm}[hbtp]
\SetAlgoLined
% \begin{algorithmic}
 \textbf{initialization: Learning rate, loss function, Epochs, and batch size}\\

 \For{\textnormal{all} Runs}
 {
    Import all expert knowledge and database\\
 }
 Split the data into the train and validation sets w.r.t \%80-\%20\\
 Define the NN based on Fig.~\ref{fig:neuralnetwork}\\
 \For{\textnormal{all} Epochs}
 {
    Import data into the model\\
    Train the model and update the weights\\
 }
 Report the accuracy and loss values for both train and validation data
 \caption{Training phase algorithm}
 \label{algo:training}
%  \end{algorithmic}
\end{algorithm}
% *****************************************Algorithm
 
To train the  neural network and find the network weights of the neurons, we need a value/loss function to find the optimal values and weights. Since we define our system model as a multi-class classification problem for the user selection, the categorical cross-entropy loss function \cite{zhang2018generalized} is suitable for this problem. The loss function is defined as:
 % *****************************************Equation
\begin{align}\label{eq:lossfunc}
\mathcal{L}(y, \hat{y}) = - \sum\limits_{z=0}^N \sum\limits_{i=0}^K (y_{zi}  \log(\hat{y}_{zi})), 
\end{align}
% *****************************************Equation
where, $N$ is the number of samples, $K$ is the number of UEs, $y$ is the true label for the UE, and $\hat{y}$ is the predicted label for the UE based on the current queue state. Afterward, we use the Adam optimization to obtain the optimal weights and minimize the loss function \cite{kingma2014adam}. 

After finalizing the model using the expert knowledge, the UAV is ready to imitate the expert as a behavioral cloner. This means, if the current state was not observed before by the expert, the UAV is likely to make a mistake. To test the UAV performance using the learned neural networks, algorithm~\ref{algo:expert} is used for the evaluation. However, in line 15 of the algorithm, instead of using the expert knowledge, the trained network can be used to predict the correct UE.

% ******************************************
% Simulation
\section{Numerical Results}\label{sec:Simulation}
To evaluate the performance, we divide this section into different parts. In the first part, we investigate the impact of the emulator parameters. The second part explains the performance of the trained deep neural network. After the training phase, we perform a comparison between the expert and the behavioral cloning over fixed arrival rates for the incoming packets. The last part investigates the case where there is a slight change in arrival rates. All simulation and emulation parts are executed on a system with AMD Ryzen 9 3900X on Ubuntu. The training phase of the behavioral cloning used Nvidia GPU RTX 2080 Ti as a resource for the computation. 

%*****************************************Figures
% \begin{figure}[bt]
% 	\centering
% 	\includegraphics[width=1\columnwidth]{Figures/arrival_distribution.pdf}
% 	\caption{Inter-arrival time for the first 30 incoming packets for five UEs. \magenta{we don't need to keep this figure}}
% % 	\vspace{-10pt}
%     \label{fig:arrival_distribution}
% \end{figure}
% *****************************************Figures

%*****************************************Figures
\begin{figure*}
    \centering
    
    \begin{subfigure}{0.23\linewidth}
        \centering
        \includegraphics[width=\textwidth]{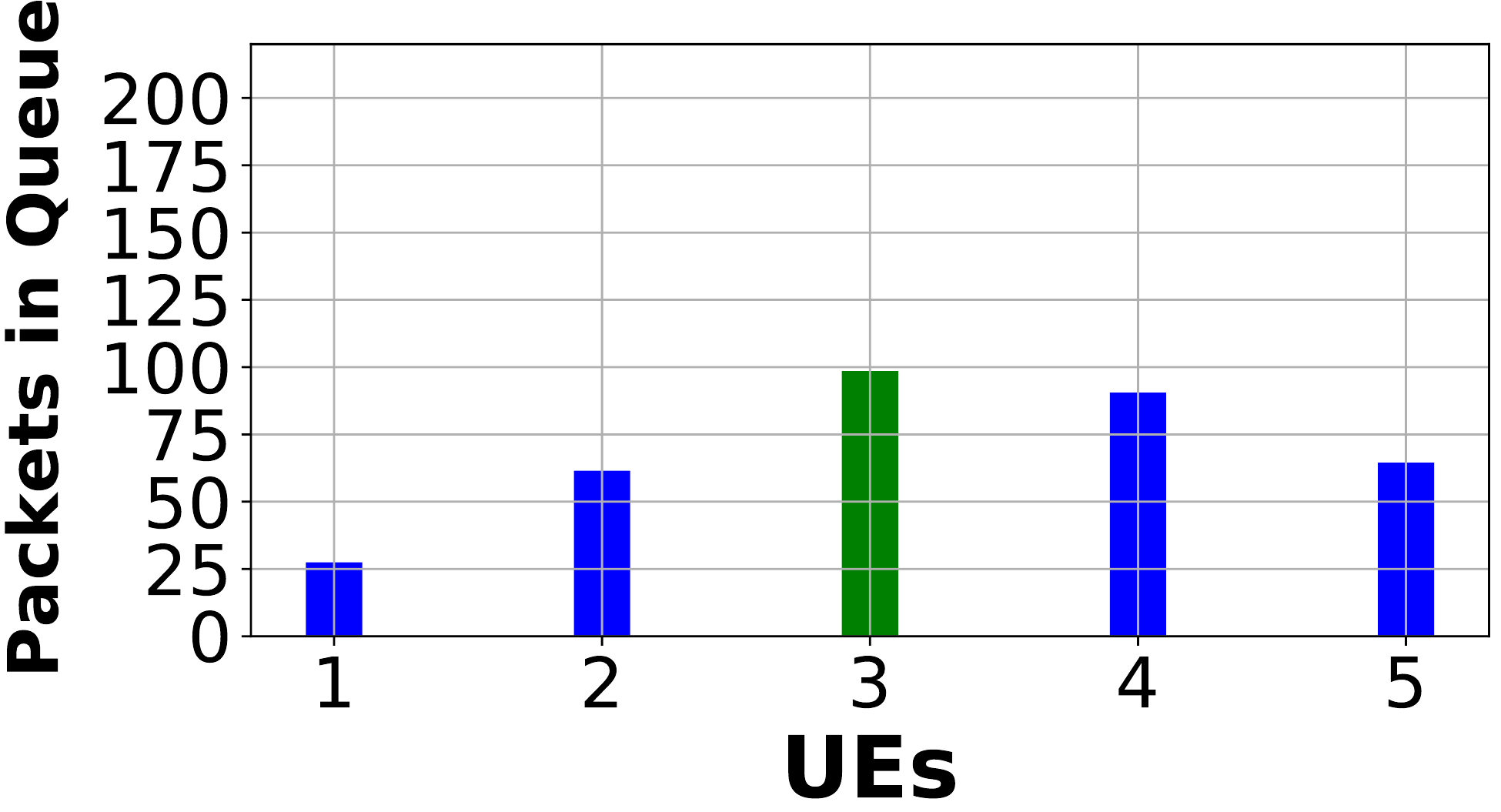}
        \caption{High priority user is UE$_3$}
        \label{subfig:snapshot_Q1}
    \end{subfigure}
    \hfill
    \begin{subfigure}{0.23\linewidth}
        \centering
        \includegraphics[width=\textwidth]{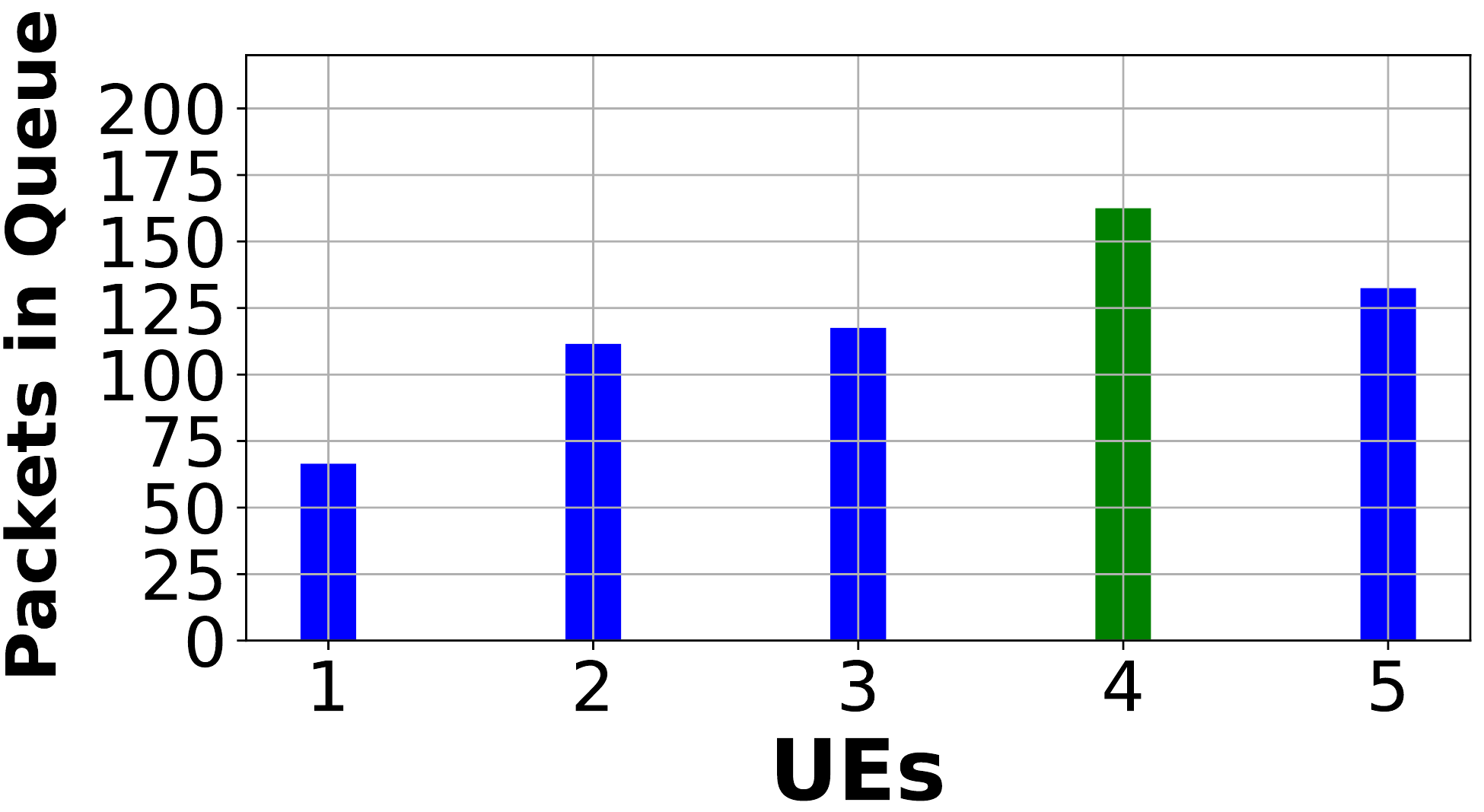}
        \caption{High priority user is UE$_4$}
        \label{subfig:snapshot_Q2}
    \end{subfigure}
    \hfill
    \begin{subfigure}{0.23\linewidth}
        \centering
        \includegraphics[width=\textwidth]{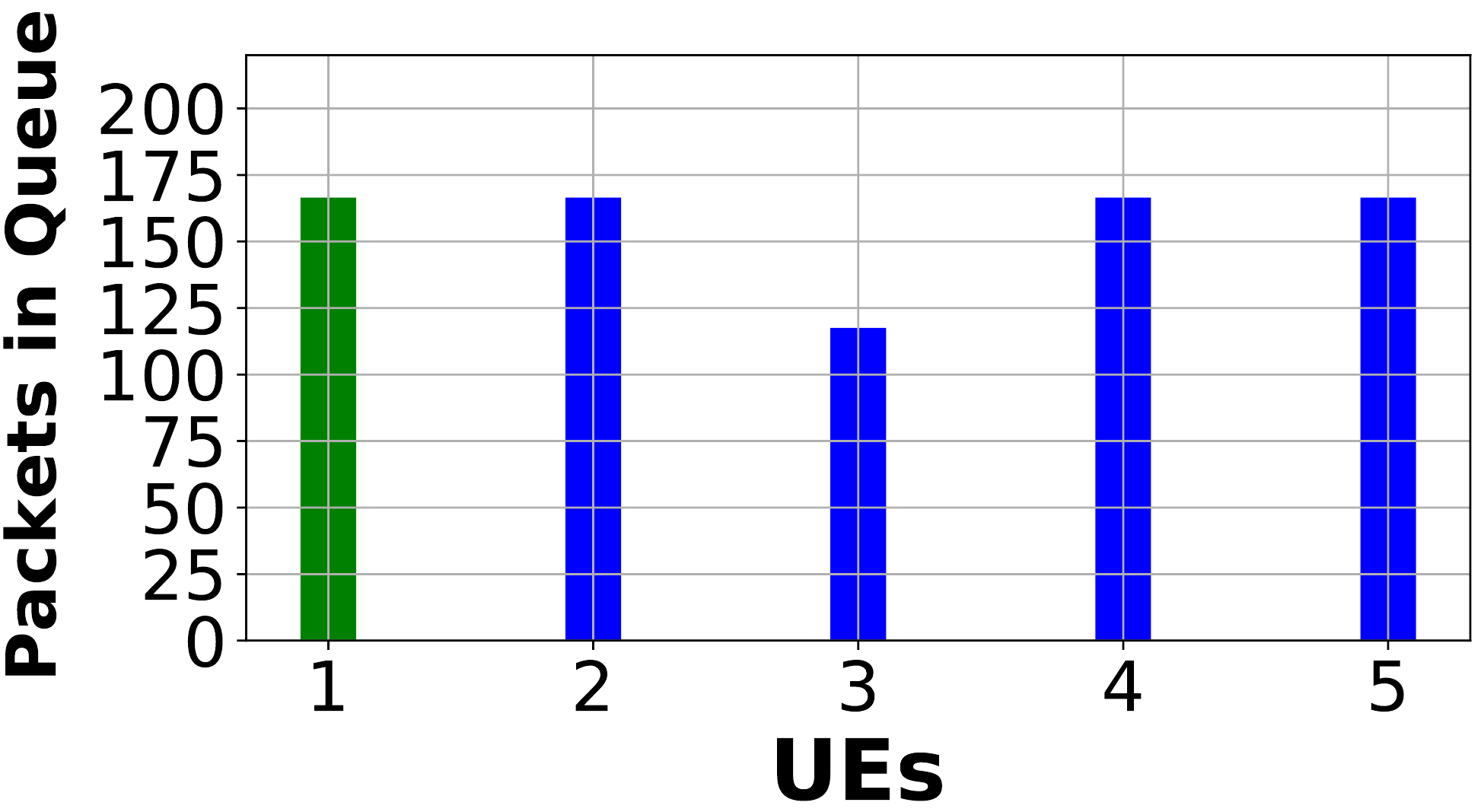}
        \caption{High priority user is UE$_1$}
        \label{subfig:snapshot_Q3}
    \end{subfigure}
    \hfill
    \begin{subfigure}{0.23\linewidth}
        \centering
        \includegraphics[width=\textwidth]{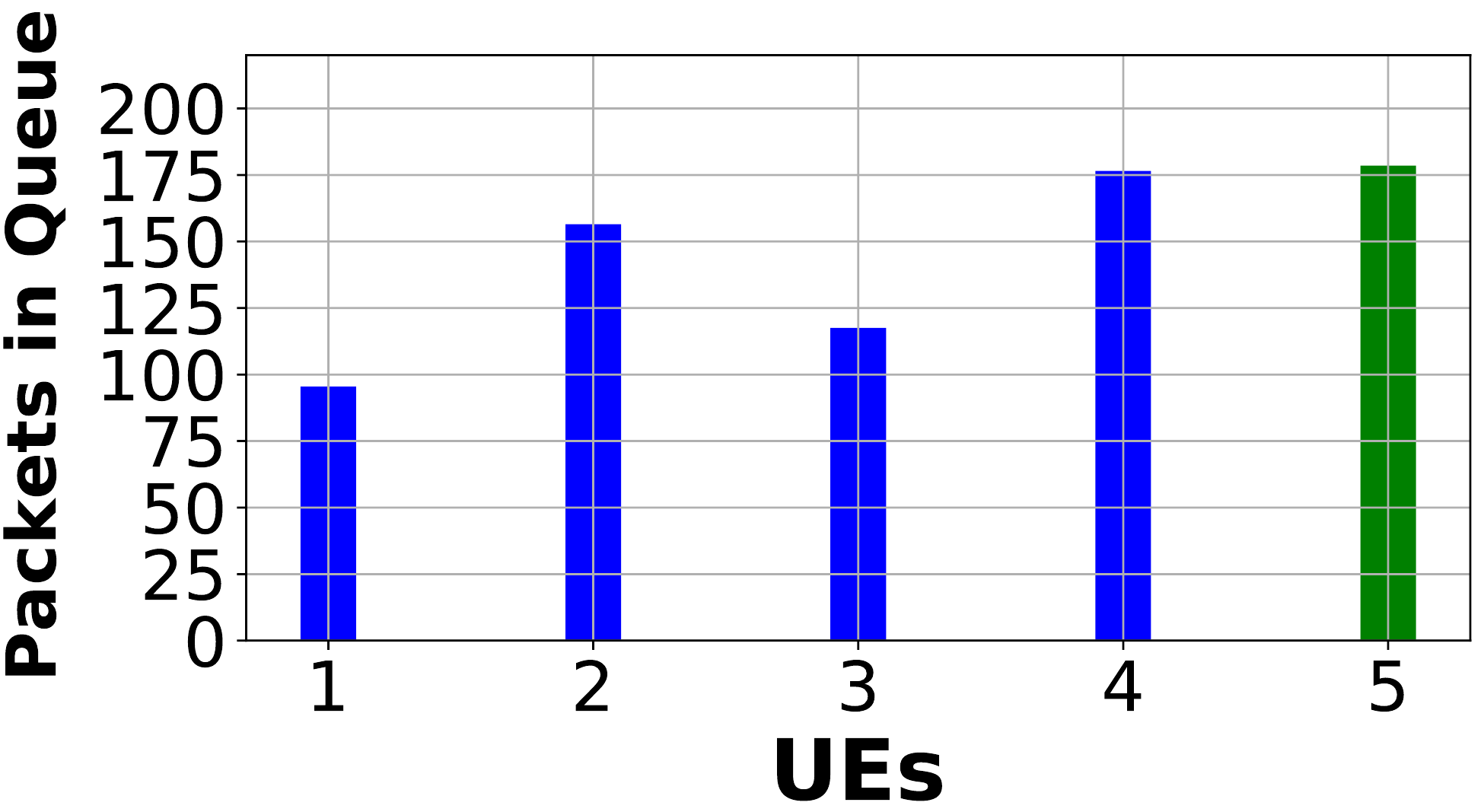}
        \caption{High priority user is UE$_5$}
        \label{subfig:snapshot_Q4}
    \end{subfigure}
    % \hfill
    % \begin{subfigure}{0.18\linewidth}
    %     \centering
    %     \includegraphics[width=\textwidth]{Figures/5_packet_cut.pdf}
    %     \caption{High priority user is UE$_2$}
    %     \label{subfig:snapshot_Q5}
    % \end{subfigure}
    
    \bigskip
    \begin{subfigure}{0.23\linewidth}
        \centering
        \includegraphics[width=\textwidth]{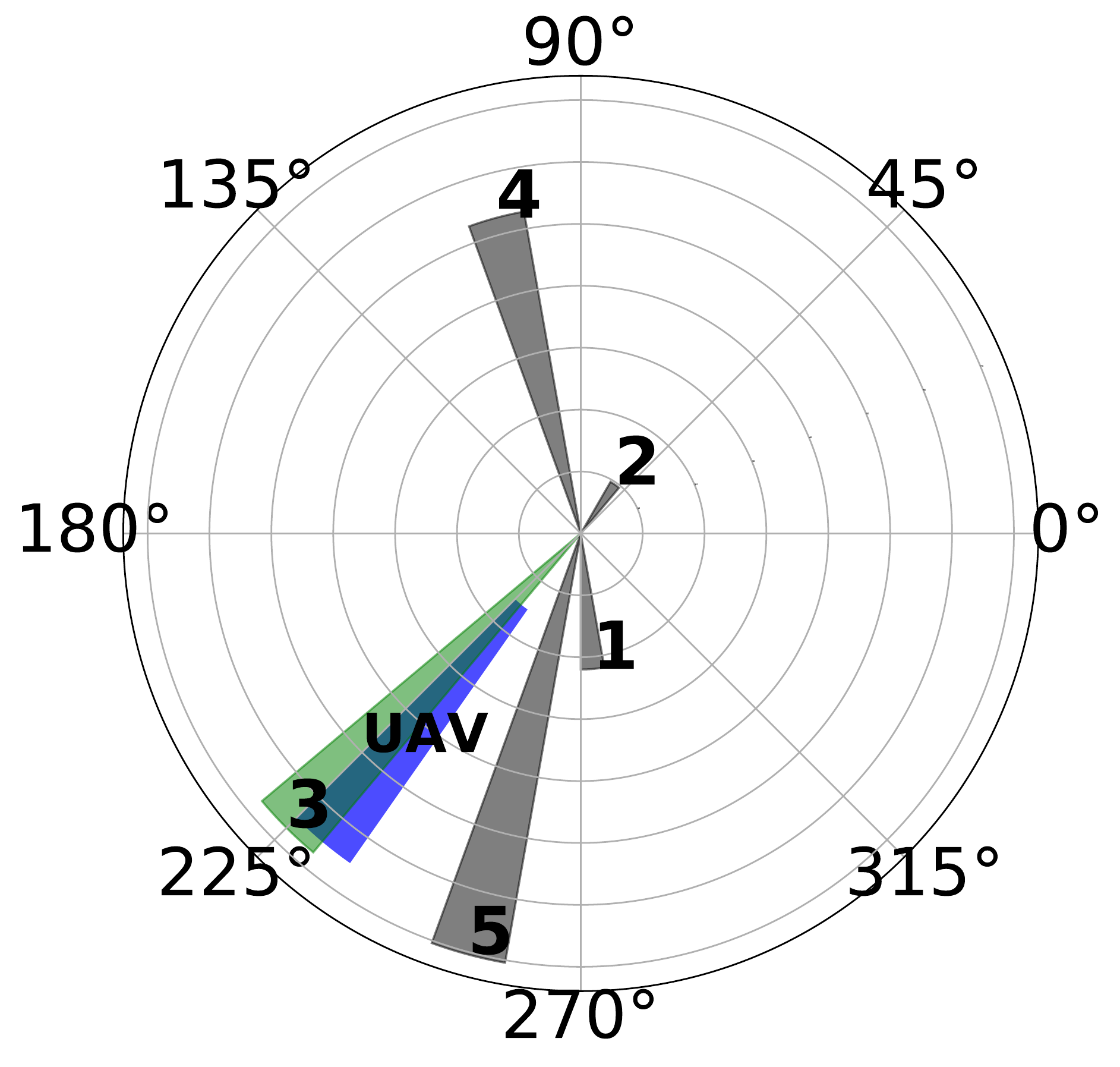}
        \caption{UAV is located in UE$_3$'s sector}
        \label{subfig:snapshot_D1}
    \end{subfigure}
    \hfill
    \begin{subfigure}{0.23\linewidth}
        \centering
        \includegraphics[width=\textwidth]{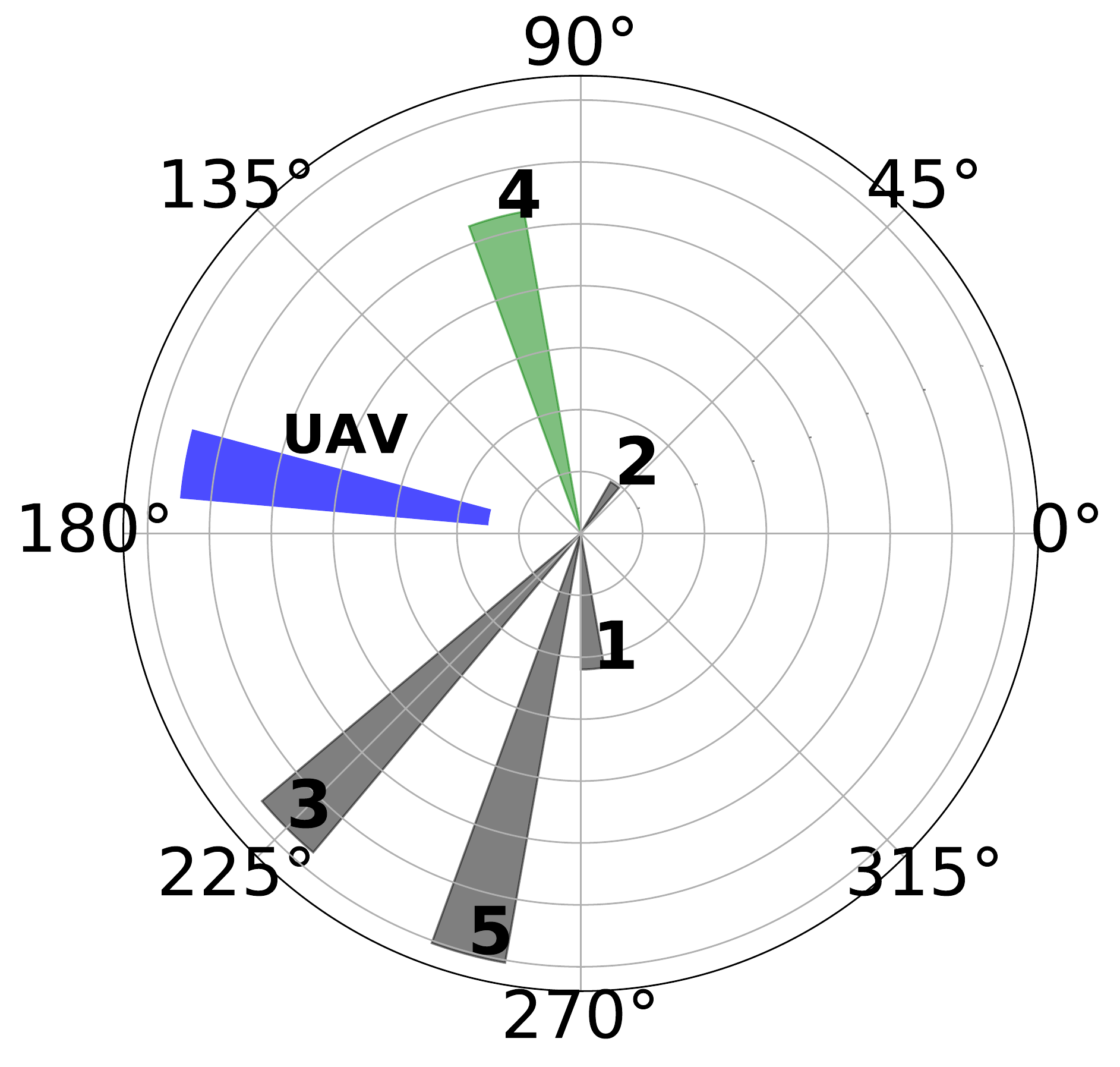}
        \caption{UAV moves toward UE$_4$}
        \label{subfig:snapshot_D2}
    \end{subfigure}
    \hfill
    \begin{subfigure}{0.23\linewidth}
        \centering
        \includegraphics[width=\textwidth]{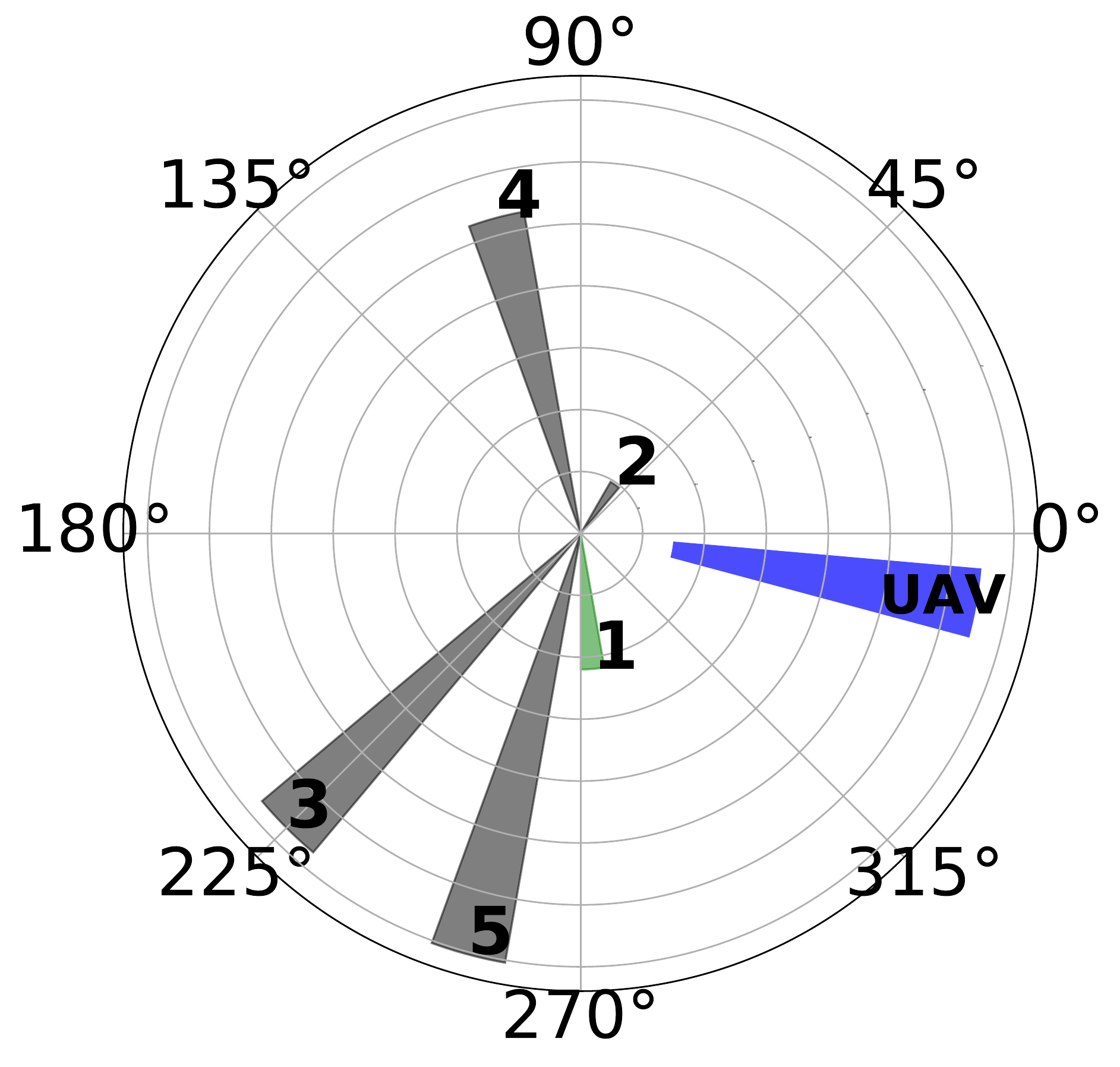}
        \caption{UAV moves toward UE$_1$}
        \label{subfig:snapshot_D3}
    \end{subfigure}
    \hfill
    \begin{subfigure}{0.23\linewidth}
        \centering
        \includegraphics[width=\textwidth]{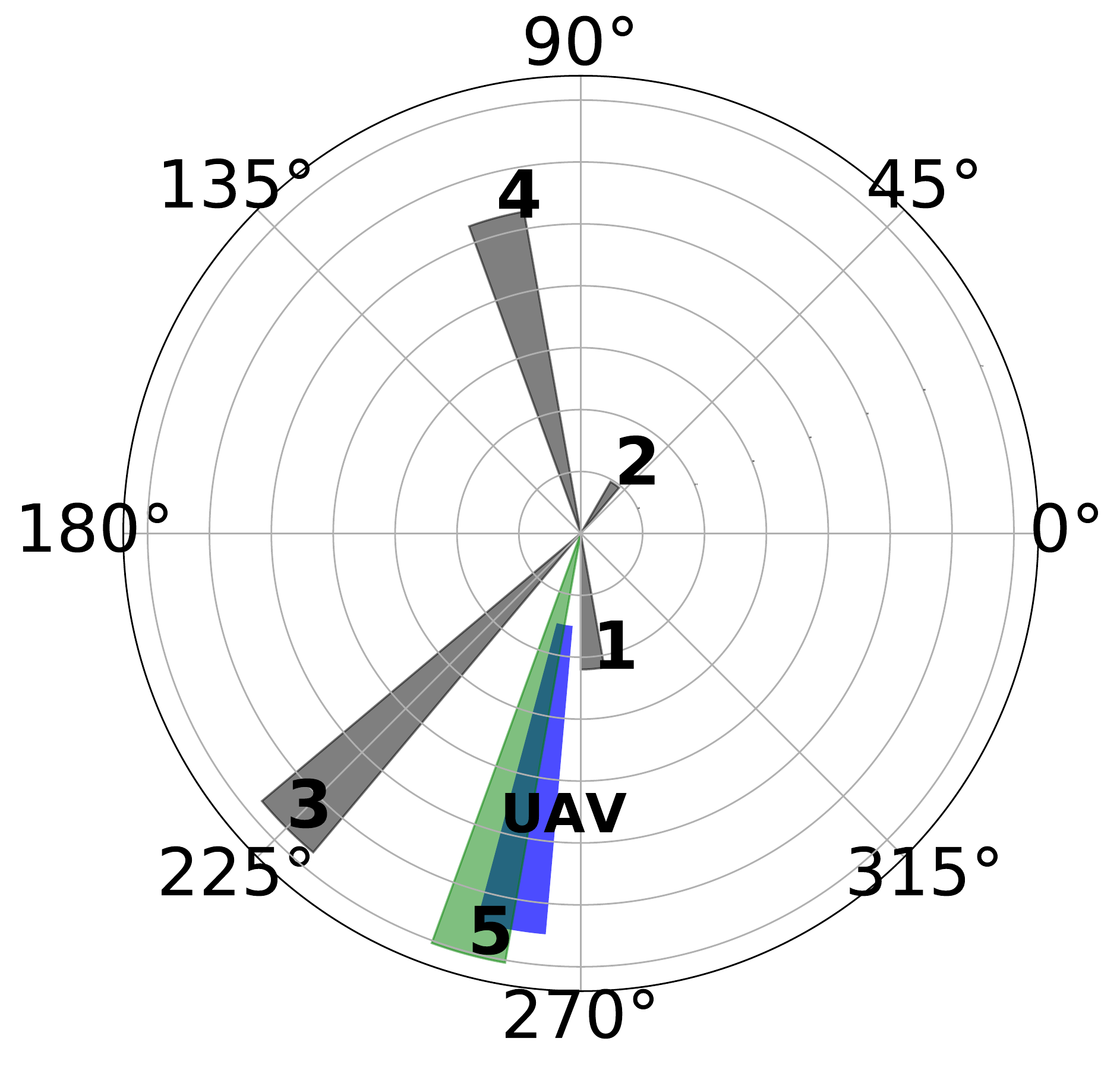}
        \caption{UAV is located in UE$_5$'s sector}
        \label{subfig:snapshot_D4}
    \end{subfigure}
    % \hfill
    % \begin{subfigure}{0.18\linewidth}
    %     \centering
    %     \includegraphics[width=\textwidth]{Figures/5_disk.pdf}
    %     \caption{UAV moves toward UE$_2$}
    %     \label{subfig:snapshot_D5}
    % \end{subfigure}
    \caption{One snapshot of the expert simulation in four different conditions}
    \label{fig:snapshot}
\end{figure*}
%*****************************************Figures

% %*****************************************Figures
% \begin{figure}[bt]
% 	\centering
% 	\begin{subfigure}[b]{1\columnwidth}
%          \centering
%          \includegraphics[width=\textwidth]{Figures/queue_emulator.pdf}
%          \caption{Queue length and dropped packet rate.}
%          \label{subfig:queue_emulator}
%      \end{subfigure}
%      \vfil
%      \begin{subfigure}[b]{1\columnwidth}
%          \centering
%          \includegraphics[width=\textwidth]{Figures/dir_emulator.pdf}
%          \caption{UAV movement in the covered area.}
%          \label{subfig:dir_emulator}
%      \end{subfigure}
%     \caption{Expert interface to monitor the queue states and the movement controller.}
%     \label{fig:emulator}
% \end{figure}
% % *****************************************Figures

\subsection{Obtaining expert knowledge using the emulator.}
In this part, we explain the emulator environment and the used parameters for the expert knowledge and experience. We assume that there is one UAV in the affected area with five UEs in the covered circle. At the beginning, the UAV is located in a random location with a random initial value for its battery energy between 40,000J and 50,000J. We also assume moving between the sectors consumes more energy compared to hovering in place. We assume that each UE has 220 packets to send in each frame which means 1100 packets are generated in each frame. That means the frame size is 220 packets for each UE. The size of each packet is 100KB and the total number of frames is 50. The expert runs the emulator for 10 times to generate enough trajectories. The covered area is considered as a 360$^\circ$ circle and divided into 36 equal sectors. Each user has a queue with a limit of 200 packets which means if the queue is full, the new incoming packets from the application layer will be dropped. The expert should take an action to choose one UE at a time to avoid packets being dropped, increase the session duration, and decrease the energy consumption. Afterward, the UAV controls its movement based on the selected UE's location to reduce the service time. We assumed that the UEs have different random arrival rates and at the beginning of each frame, the packet arrival time is generated based on a random Poisson distribution with those fixed arrival rates $\lambda = (\lambda_1, \lambda_2, \dots, \lambda_5)$ = (3, 5, 10, 8, 7). Moreover, the service time for each user is calculated based on a random Exponential distribution with a fixed rate and we assumed that all the users have a fixed rate, $\mu = (\mu_s, \mu_s, \dots, \mu_s)$.
% Figure~\ref{fig:arrival_distribution} shows the distribution of inter-arrival times for five UEs in one single frame. For the sake of the demonstration, we only showed the first 30 packets of each UE.
Also, a snapshot of the emulator for the queue lengths and UAV movement control system is shown in Fig.~\ref{fig:snapshot}. The first row shows the status of buffers for all UEs. The second row shows the UAV's controlling system regarding its movement. The green bar chart shows the high priority UE which the UAV is servicing at that moment. 
% The red bar charts demonstrate the accumulative rate of the dropped packets.
In the controlling system, the high priority UE is shown with a green sector and the UAV is highlighted with a blue one. From Figures~\ref{subfig:snapshot_Q1} and \ref{subfig:snapshot_D1} to Figures~\ref{subfig:snapshot_Q4} and \ref{subfig:snapshot_D4}, the UAV selects different UEs and changes its location accordingly.
The full demonstration of this interaction between the expert and the UEs is shown in \cite{youtube2020_imitation}. At the end of this phase, 100,000 seconds $\sim$ 28 hours of the expert interaction for 550,000 packets is stored for the training phase. To obtain and collect the expert knowledge, an emulator is set up in Python 3.6. The emulator uses a single drone at this version, but it has the capability to develop it for multiple UAVs at the same time. The altitude of the drone is a variable that the user can change. Also, the dimension of the emulator and coverage area radius are other variables that can be configured in the code. The code for the emulator is available in \cite{github:code_IL_BC2020}.

\subsection{Training the deep neural network}
In Deep Learning (DL) training, the performance of the trained behavioral cloning model supports imitation. To train the model, all gathered data from the expert is split into two sets of training and validation  w.r.t 80\%-20\%. The initial learning rate for the Adam optimizer is $0.001$ and the decay rate for the learning rate is equal to initial rate over the number of epochs. 40 epochs are considered for the training phase.

\begin{figure}[bt]
	\centering
	\includegraphics[width=1\columnwidth]{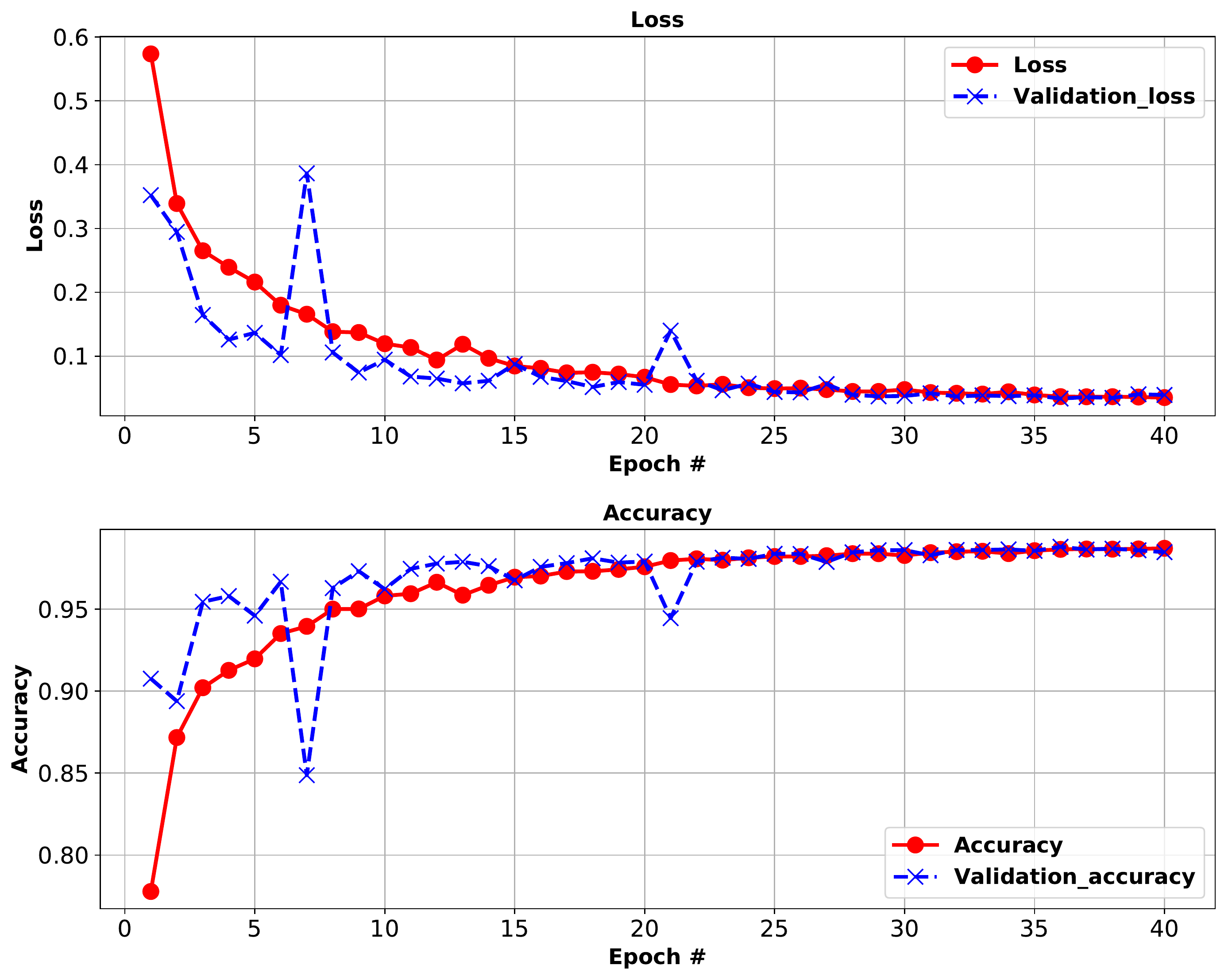}
	\caption{Accuracy and loss values for the training data over 20 epochs for user selection based on queue states.}
% 	\vspace{-10pt}
    \label{fig:accuracy_loss_training_user}
\end{figure}

In Fig.~\ref{fig:accuracy_loss_training_user}, the accuracy and loss of the trained model for the training and validation dataset over 20 epochs is shown for the user selection based on the queue states of UEs. The Figure~\ref{fig:accuracy_loss_training_user}-top shows the loss data and Figure~\ref{fig:accuracy_loss_training_user}-bottom shows the accuracy metric. Table~\ref{tab:accuracy_loss} reports the accuracy and loss for three different datasets: 1) Training set, 2) Validation set, and 3) Test set. The training and validation set were used to train the model and find the weights. However, the test set was never used in the training phase and it is totally new to the model.

%*****************************************Figures
% *****************************************Figures

% *****************************************Table
\begin{table}[bth!]
\caption{Accuracy and loss for training, validation, and test sets .}
 \centering{
\label{tab:accuracy_loss}
\resizebox{0.68\linewidth}{!}{  %fit to windows command
% \color{blue}
\begin{tabular}{c|c|c}
\toprule
\toprule
\Topspace
\Bottomspace
\textbf{\Large} & \multicolumn{2}{c}{\textbf{\Large Performance}}
\\
\hline
\toprule
\toprule
\Topspace
\Bottomspace
\textbf{\large Dataset} & {\large Loss} & {\large Accuracy(\%)}
\\
\hline
\Topspace
\Bottomspace
{\fontsize{14}{17}\selectfont Test set} & \textbf{\fontsize{14}{17}\selectfont $0.09370$} & \textbf{\fontsize{14}{17}\selectfont $97.52$}
\\
\hline
\Topspace
\Bottomspace
{\fontsize{14}{17}\selectfont Validation set} & \fontsize{14}{17}\selectfont $0.0391$ & \fontsize{14}{17}\selectfont $98.47$
\\
\hline
\Topspace
\Bottomspace
{\fontsize{14}{17}\selectfont Training set} & \fontsize{14}{17}\selectfont $0.0350$ & \fontsize{14}{17}\selectfont $98.71$
\\
\bottomrule 
\end{tabular}
} % end of \resizebox{1\linewidth}{!}{
} % end of \centering{
% \vspace{-10pt}
\end{table}
% *****************************************Table

\begin{figure}[bt]
	\centering
	\includegraphics[width=1\columnwidth]{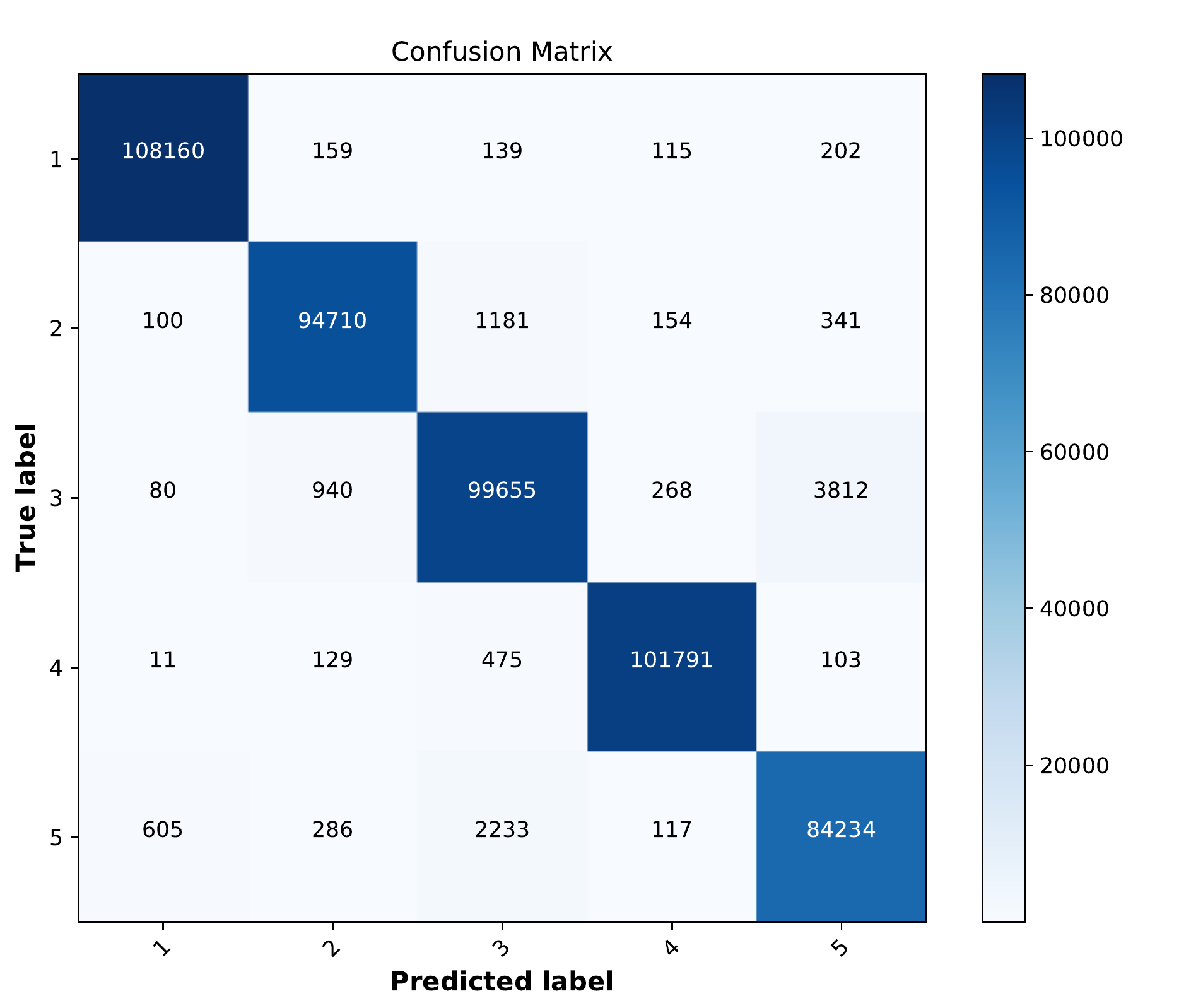}
	\caption{Confusion matrix for the true and predicted UE in 50 frames of data transmission.} 
% 	\vspace{-10pt}
    \label{fig:confusion_mat}
\end{figure}

In addition, Figure~\ref{fig:confusion_mat} illustrates the confusion matrix plot for the UE prediction and selection compared with the true labels in the test scenario. This matrix is plotted for five UEs to report a summary how accurate the behavioral cloning is performing given all queue states. Training the imitated model on 28 hours of expert's trajectories (set of states and actions) only takes less than half an hour. Later this trained model can be used in real-time scenarios on the UAV's computer. Nvidia Jetson Nano \cite{NVIDIAJe4:online} is a good example of a mini-computer equipped with Nvidia GPU capable of running light on-board Tensorflow. The time complexity of the decision making does not depend on the training data and it only depends on the mini-computer performance since the imitated model has been already trained before. The UnVAIL model is imported to the Jetson Nano to test the time complexity. Jetson returns the chosen UE based on the imitated expert policy in less than a millisecond. It is worth noting that the input data for the Jetson was not based on a real-flight or an LTE network, it was a dataset generated by another computer.

%*****************************************Figures

\subsection{IL-Behavioral cloning performance}
\begin{figure}[bt]
	\centering
	\includegraphics[width=1\columnwidth]{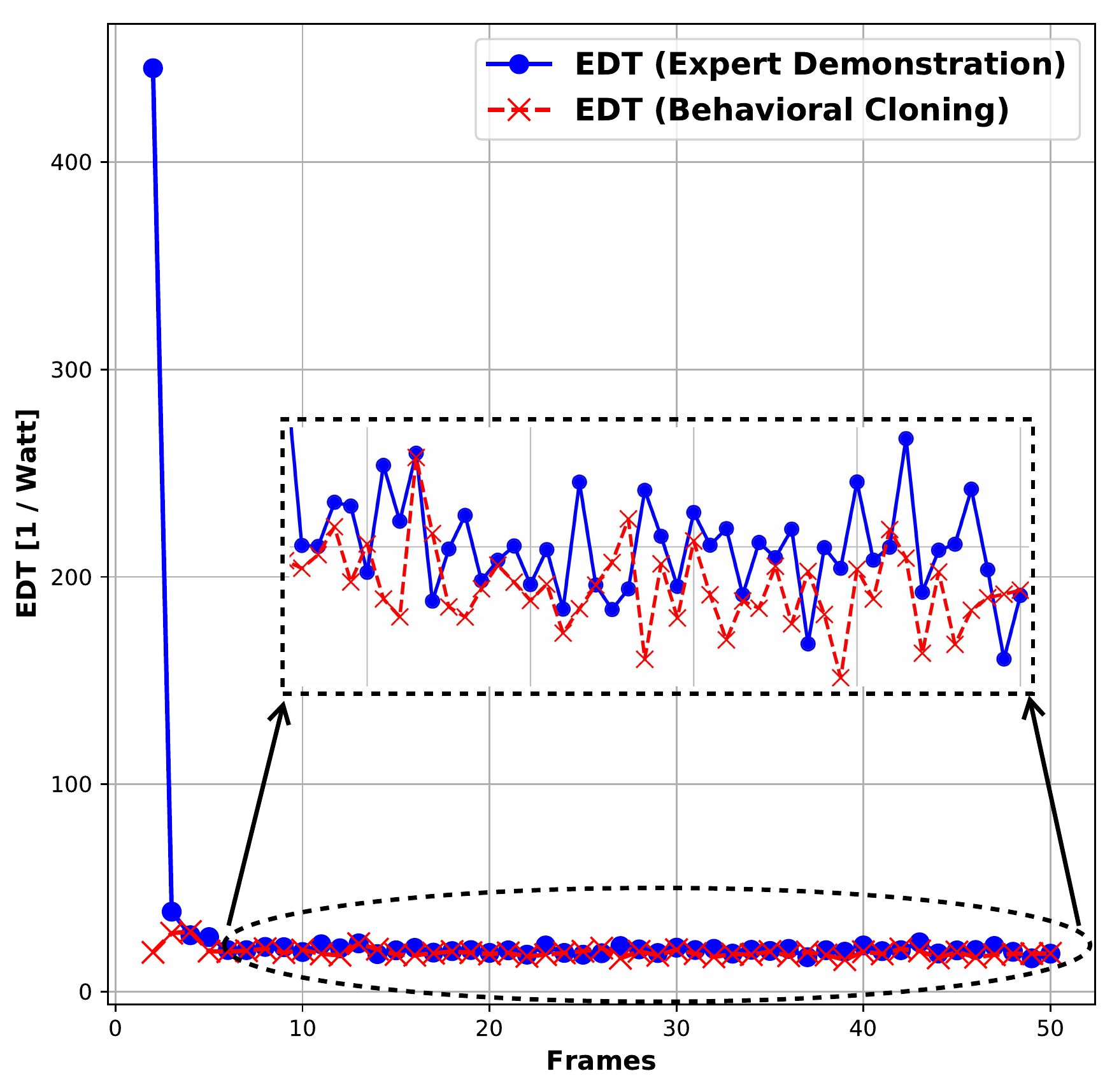}
	\caption{Energy delay throughput for both the expert and the imitated model by the UAV.} 
% 	\vspace{-10pt}
    \label{fig:edt_expert_UAV}
\end{figure}

The goal of the evaluation is to determine how well the UAV imitates the expert. This section does not intend to investigate the optimality of the solution, since it is assumed that the expert's demonstration is already optimal. Two scenarios were conducted with the exact same inter-arrival rates for the incoming packets for all UEs. However, in both cases, the inter-arrival times are generated randomly based on the constant rates and the Poisson distribution. Figure~\ref{fig:edt_expert_UAV} illustrates the energy-delay throughput (EDT) versus all frames for both the expert and the UAV. The EDT was defined in (\ref{eq:util}) in Sec.~\ref{sec:System_Model}. At the beginning of the session, the EDT value is higher for both of the UAV and the expert. Based on the observation, it can be concluded that the UAV mimics the expert behavior with a reasonable performance. 

%*****************************************Figures
% *****************************************Figures

\begin{figure}[bt]
	\centering
	\includegraphics[width=1\columnwidth]{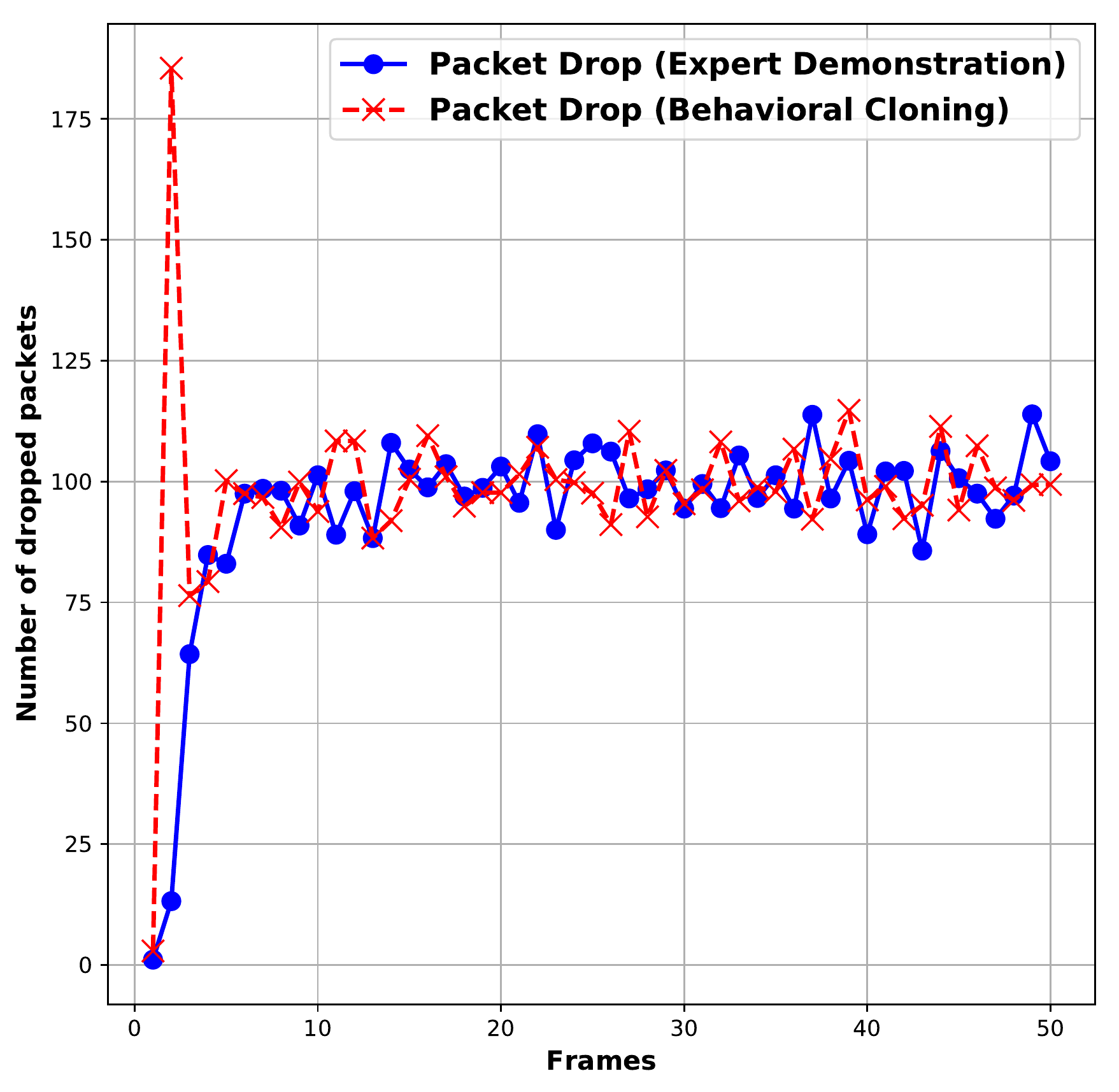}
	\caption{Number of dropped packets for both the expert and the autonomous UAV using the imitation learning.} 
% 	\vspace{-10pt}
    \label{fig:drop_expert_UAV}
\end{figure}

Figure~\ref{fig:drop_expert_UAV} shows the number of dropped packets versus 50 frames for all UEs for both the expert and the UAV. Each point on the plot is a summation of dropped packets for all UEs over all events in one frame. At  early frames, the packet dropping rate is lower since the queues are less occupied at the beginning of the session and the probability of packet dropping is lower. Increasing the number of UEs with the same rates of packet arrival and service time will increase the number of dropped packets compared to Figure~\ref{fig:drop_expert_UAV}. However, in general, the number of dropped packets depends on the number of UEs and packet arrival and service time rates and it is a trade-off between these values.

%*****************************************Figures
% *****************************************Figures

%\magenta{can you add another simulation scenario to show the selected UE by the expert and the UE, and probably the trajectory} \brown{(I can add that my only concern is it will take two much space for both showing the UE's buffer and distance between UAV in four different examples. And we have the video, it is going to show the same concept of demonstration.)} \yellow{is there any page limit? can help if the reviewers do not watch the video, which they probably don't}

\begin{figure}[bt]
	\centering
	\includegraphics[width=1\columnwidth]{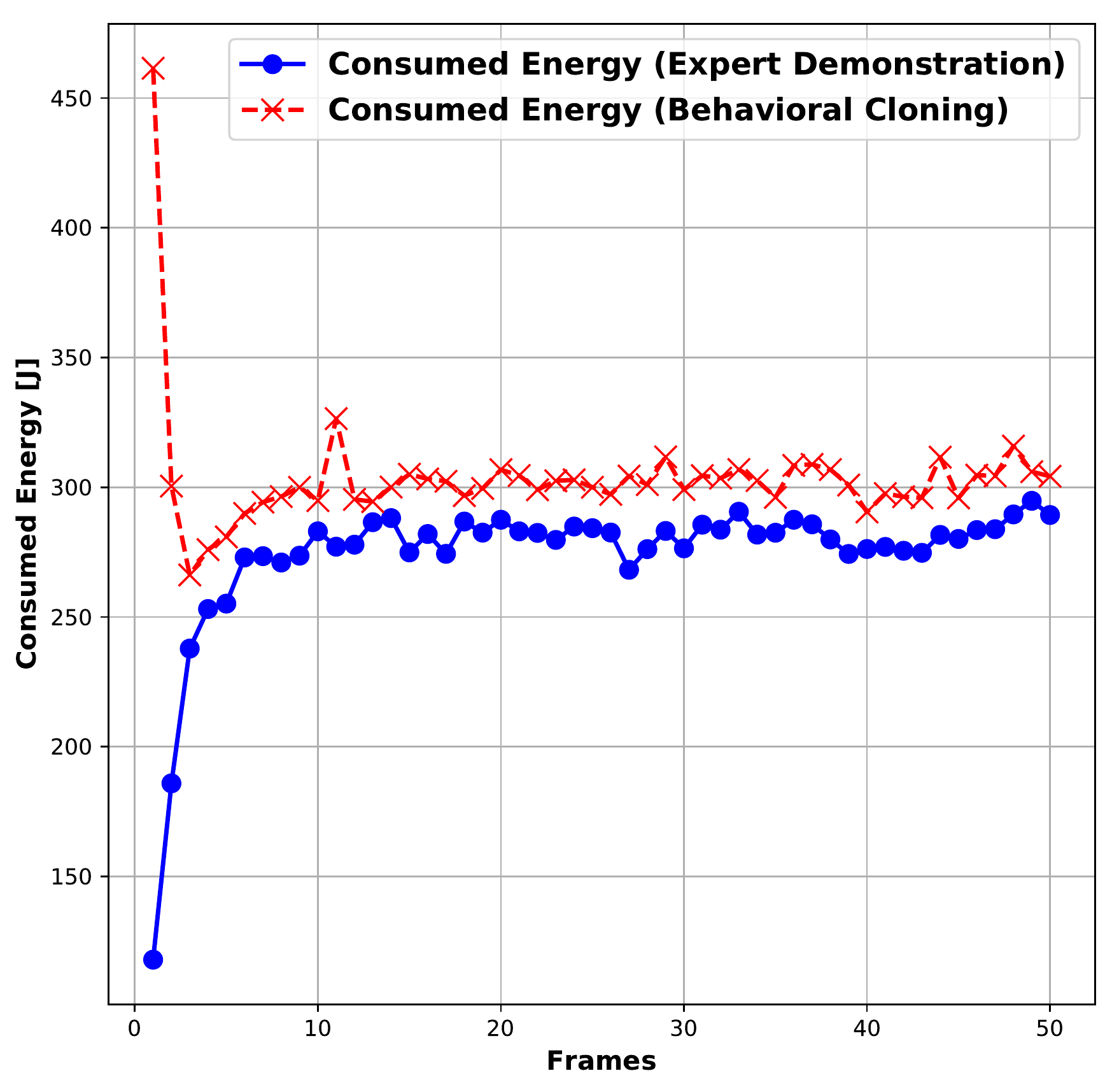}
	\caption{Consumed energy based on hovering, movement, and transmission for both the expert the imitated model.} 
% 	\vspace{-10pt}
    \label{fig:energy_expert_UAV}
\end{figure}

Figure~\ref{fig:energy_expert_UAV} shows the total energy consumption rate for the drone versus all frames in both scenarios of the expert and the behavioral cloning. Each point in the graph is the summation of packet transmission, and the mobility energy consumption rates. There are two reasons that the behavioral cloning method has a higher energy consumption rate as compared to the expert policy. (1) the inter-arrival rates for the packets are not exactly the same. Although they have the same mean rates in both cases, they are generated completely at random, (2) in behavioral cloning, the agent (UAV) cannot realize the full policy of the expert or the reward function; and instead it tries to mimic the expert without any fundamental model of the scenario. And since the behavioral cloning in this paper is a model-free approach, the learner cannot reconstruct the reward function in (\ref{eq:util}). As a result, it switches more often to mimic the expert and this increases the energy consumption rate. 
%*****************************************Figures
% *****************************************Figures

\begin{figure}[bt]
	\centering
	\includegraphics[width=1\columnwidth]{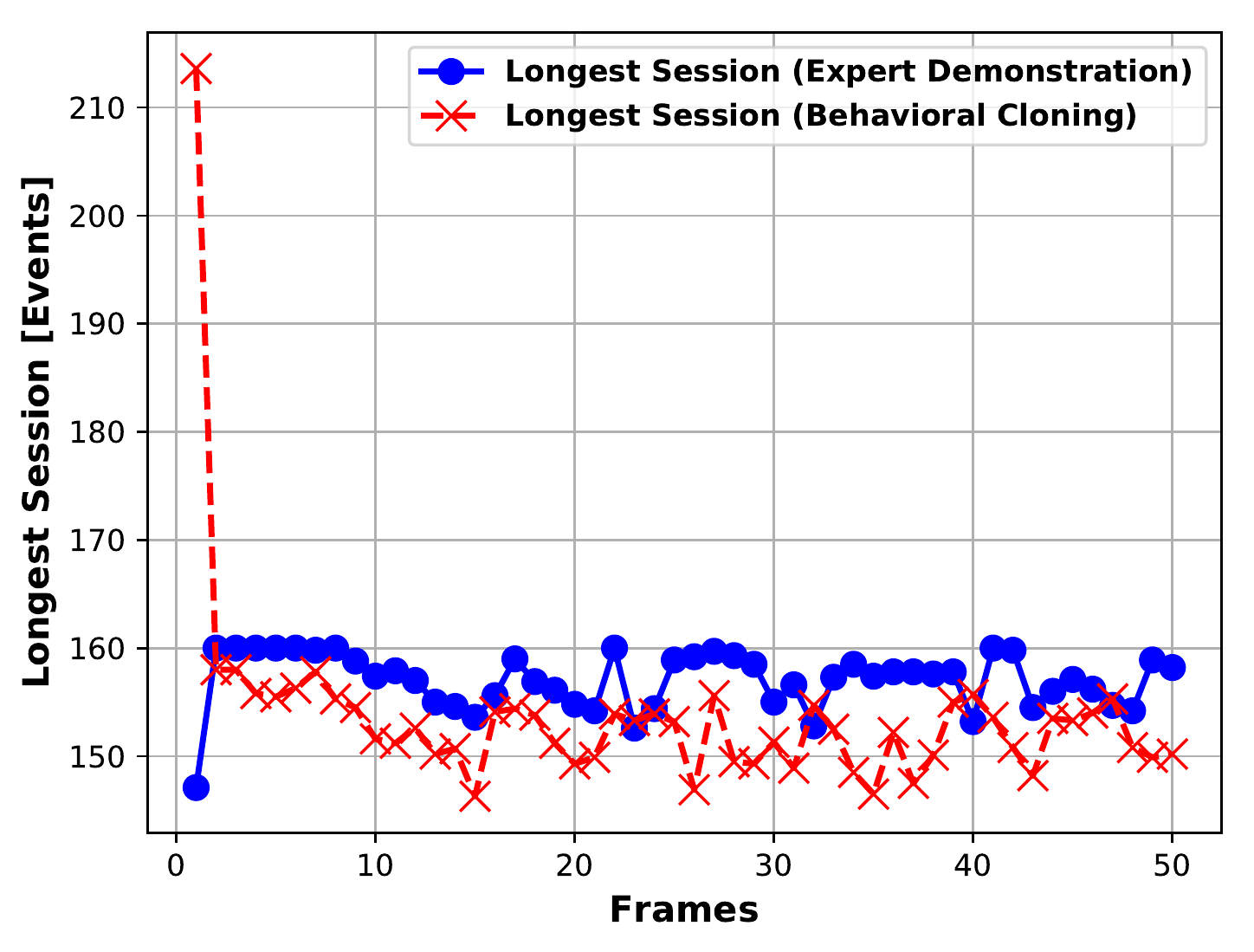}
	\caption{Longest session based on number of events or transmission for both the expert and the UAV.} 
% 	\vspace{-10pt}
    \label{fig:session_expert_UAV}
\end{figure}

Figure~\ref{fig:session_expert_UAV} reports the longest session versus all 50 frames. In each frame, 1000 events occurred. Each point in the graph shows the longest session for one UE that the UAV keeps scheduling its packet to the base station before switching to another UE. It is assumed that 1000 events are the total number of events in each frame. Since the imitation model tries to mimic the expert regarding the queue states, it could not completely realize the intention for having a long session communication. The longest session for all 50 frames is recorded as 160 transmissions with no interruption for the expert and 158 transmissions for the behavioral cloning.

%*****************************************Figures
% *****************************************Figures

%*****************************************Figures
\begin{figure}[bt]
	\centering
	\includegraphics[width=1\columnwidth]{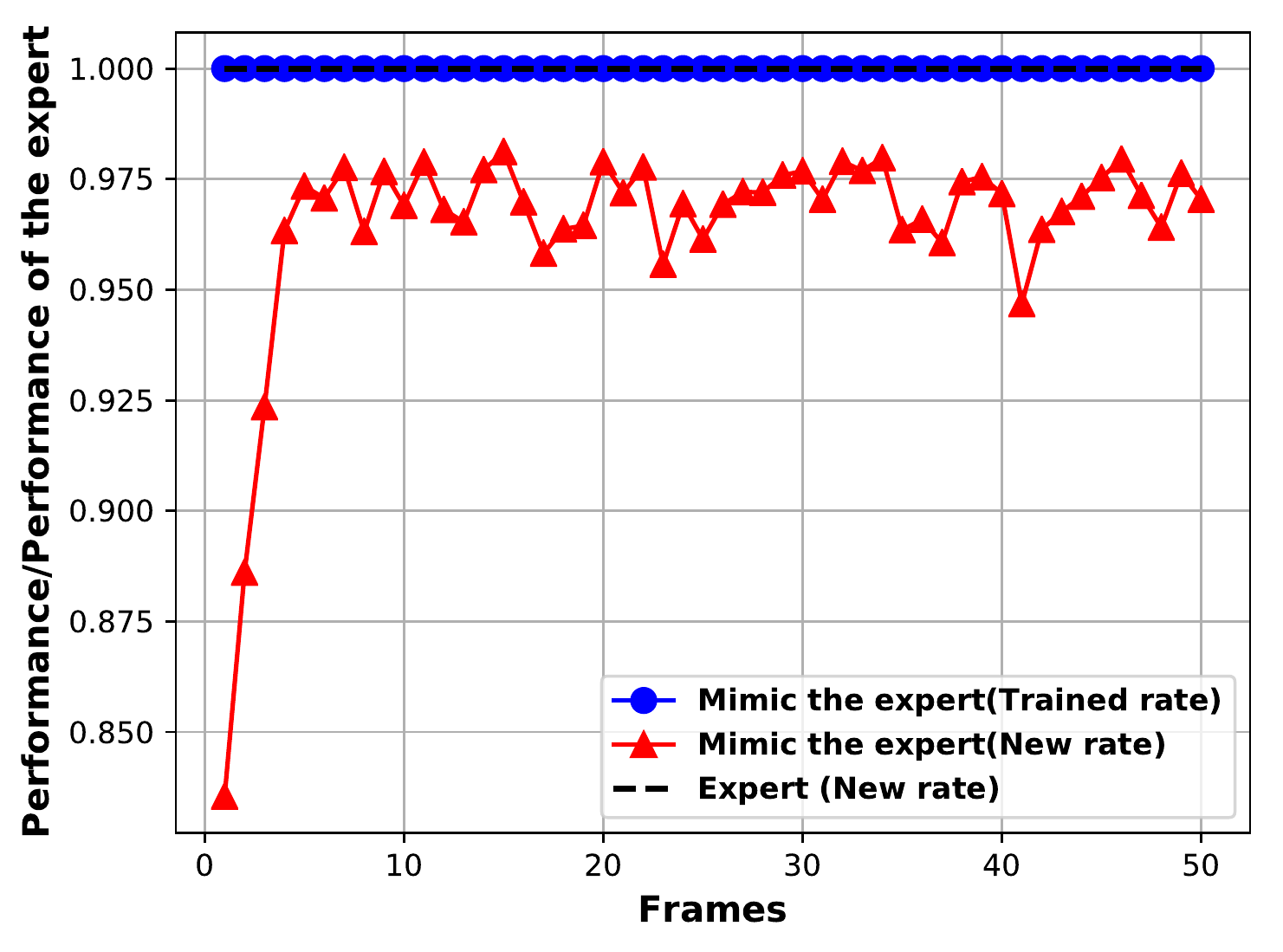}
	\caption{Performance plot for the expert and the imitation learning with a new inter-arrival rates for UEs.} 
% 	\vspace{-10pt}
    \label{fig:newrate_performance}
\end{figure}
% *****************************************Figures

\subsection{Performance comparison for the new arrival rates}
In this part, the packet inter-arrival rates is slightly changed for the UEs. The purpose is to investigate the robustness of the imitated model based on the previous inter-arrival rates for the current changes in the environment. Here, another scenario with both the expert knowledge and the behavioral cloning model is analyzed. Figure~\ref{fig:newrate_performance} shows the performance of the UAV using the behavioral cloning and compares it with the expert performance. It can be inferred that, those small changes in the inter-arrival rates brought new queue states which had never been observed before in the training phase. Hence, those states were not trained in the behavioral cloning model. As a result, one can observe that the performance is degraded compared to the expert performance.
The average accuracy of the behavioral cloning in this scenario for 50 frames is 79.436\% and the reported loss is 20.7759 which are a huge change compared to Table~\ref{tab:accuracy_loss}. To compensate the accuracy and loss regarding those small changes in the state-action trajectories and address the related issue, we aim at implementing similar approaches for the expert-knowledge demonstrations such as DAgger, RL, and IRL to compare the performance with UnVAIL and other methods in the future.

% ******************************************
% Future works and open challenges
% \section{Discussion}
% \label{sec:Discussion}

% ******************************************
% Conclusion
\section{Conclusion}
\label{sec:Conclusion}

This paper introduces a model-free approach for a UAV-assisted communication in a sparsely populated remote area where a natural or man-made disasters such as flood or wildfire has completely damaged the base stations. In the developed UnVAIL (UAV-Assisted Imitation Learning) approach, the user equipment (UEs) have limited buffers to store the queued packets for the transmission. The packets arrival time is randomly generated based on a Poisson process with fixed arrival rates. The proposed UnVAIL method utilizes a data-driven learning method called imitation learning (behavioral cloning) to train the UAV using a deep neural network based on a human expert knowledge and trajectories using a simulator. The UAV's strategy is to select an UE at a proper time to minimize the packet dropping rate, prolong the UAV's battery lifetime, and minimize the number of switches between different UEs. The simulation results show that the UAV mimicked the expert behavior with approximately 97\% accuracy, likewise the comparative energy throughput overt the the number of dropped packets, the consumed energy, and the longest session between the expert and the learner UAV. The proposed IL-based approach is evaluated in the scenarios which have not been seen in the demonstration by using different arrival rates to show how well the agent can determine an optimal strategy for these unseen scenarios using the previously trained model. Future steps include leveraging modeling and tools to assist in situation awareness.

% In this paper, we introduced a model-free approach for a UAV-assisted communication in a sparse remote area where a natural or man-made disaster such as flood or wildfire has completely damaged the base station. In our approach, the user equipment (UEs) have limited buffers to store the queued packets for the transmission. The packets arrival time is randomly generated based on a Poisson process with fixed arrival rates. The proposed method utilized a data-driven learning method called imitation learning (behavioral cloning) to train the UAV using a deep neural network based on a human expert knowledge and trajectories using a simulator. The UAV's strategy is to select an UE at a proper time to minimize the packet dropping rate, prolong the UAV's battery lifetime, and minimize the number of switches between different UEs. In the simulation results, we showed that the UAV mimicked the expert behavior with approximately 97\% accuracy. We compared the energy throughput, the number of dropped packets, the consumed energy, and the longest session between the expert and the learner UAV. We also evaluated the performance of the proposed IL-based approach for the scenarios not seen in the demonstration by using different arrival rates and used the previous trained model to show how well the agent can determine an optimal strategy for these unseen scenarios.

% \section*{Appendix}

% Appendixes, if needed, appear before the acknowledgment.

% \section*{Acknowledgment}

\section{Acknowledgement}
This material is based upon work supported by the Air Force Office of Scientific Research under award number FA9550-20-1-0090 and the National Science Foundation under Grant Numbers CNS-2034218, CNS-2039026, and ECCS-2030047. Any opinions, findings and conclusions or recommendations expressed in this material are those of the author(s) and do not necessarily reflect the views of the US government or AFRL. Distribution A: Approved for Public Release, distribution unlimited. Case Number AFRL-2021-1039 on March 30, 2021.

% The preferred spelling of the word ``acknowledgment'' in American English is 
% without an ``e'' after the ``g.'' Use the singular heading even if you have 
% many acknowledgments. Avoid expressions such as ``One of us (S.B.A.) would 
% like to thank $\ldots$ .'' Instead, write ``F. A. Author thanks $\ldots$ .'' 
% \textbf{Sponsor and financial support acknowledgments are placed in the 
% unnumbered footnote on the first page, not here.}

\bibliography{main}
\bibliographystyle{IEEEtran}

\begin{IEEEbiographynophoto}{A. Shamsoshoara} received a B.Sc. degree in Electrical Engineering and an M.Sc. degree in Electrical and Communication Engineering in 2012 and 2014, respectively. Also, he received an M.Sc. degree in Informatics from Northern Arizona University in 2018. He worked as a network engineer from 2015 to 2017. Currently, he is a Ph.D. candidate in the School of Informatics, Computing \& Cyber Systems at Northern Arizona University. His main research interests are wireless networks, UAV networks, spectrum sharing, and machine learning.
\end{IEEEbiographynophoto}

\begin{IEEEbiographynophoto}{F. Afghah} is an Associate Professor with the School of Informatics, Computing and Cyber Systems, Northern Arizona University, where she is the Director of Wireless Networking and Information Processing Laboratory. Her research interests include wireless communication networks, decision making in multi-agent systems, radio spectrum management, and AI-assisted healthcare. Her research has been continually supported by NSF, AFRL, AFOSR and ABOR. She is the recipient of several awards including the AFOSR Young Investigator Award (2019), NSF CAREER Award (2020), NAU's Most Promising New Scholar Award (2020), and NSF CRII Award (2017). She is the author/co-author of over 90 peer-reviewed publications.
\end{IEEEbiographynophoto}

\begin{IEEEbiographynophoto}{E. Blasch} 
received the B.S. degree in mechanical engineering from MIT in 1992, the master’s degrees in mechanical, health science, and industrial engineering (human factors) from Georgia Tech in 1994, 1995, and 1995, respectively, the M.B.A. degree in 1998, the M.S.E.E. degree in 1998, the M.S. Econ degree in 1999, and the Ph.D. degree in electrical engineering in 1999 from Wright State University. He is a graduate of Air War College in 2008. He is also a Program Manager for the AFOSR DDDAS Program. In 1996, he served in Active Duty with the United States Air Force. He is a Principal Scientist with the U.S. Air Force Research Lab, Information Directorate, Rome, NY, USA.
\end{IEEEbiographynophoto}

\begin{IEEEbiographynophoto}{J. Ashdown} was born in Niskayuna, NY, USA, in 1984. He received the B.S., M.S., and Ph.D. degrees from Rensselaer Polytechnic Institute, Troy, NY, USA, in 2006, 2008, and 2012, respectively, all in electrical engineering. His Ph.D. dissertation was on a high-rate ultrasonic through-wall communication system using MIMO-OFDM in conjunction with interference mitigation techniques. In 2012, he was a recipient of the Best Unclassified Paper Award at the IEEE Military Communications Conference. From 2012 to 2015, he worked as a Civilian Research Scientist with the Department of Defense (DoD), SPAWAR Systems Center Atlantic, Charleston, SC, USA where he was involved in several basic and applied research projects for the U.S. Navy, mainly in the area of software defined radio. In 2015, he transferred within DoD. He is currently an electronics engineer with Air Force Research Laboratory, Rome, NY, USA, where he is involved in the research and development of advanced emerging technologies for the U.S. Air Force.
\end{IEEEbiographynophoto}

\begin{IEEEbiographynophoto}{E. Bennis} 
is currently an Associate Professor with the Centre for Wireless Communications, University of Oulu, Oulu, Finland, where he is also an Academy of Finland Research Fellow and the Head of the Intelligent Connectivity and Networks/Systems Group (ICON). He has coauthored one book and published more than 200 research articles in international conferences, journals, and book chapters. His current research interests include radio resource management, heterogeneous networks, game theory, and machine learning in 5G networks and beyond. Dr. Bennis was a recipient of several prestigious awards, including the 2015 Fred W. Ellersick Prize from the IEEE Communications Society, the 2016 Best Tutorial Prize from the IEEE Communications Society, the 2017 EURASIP Best Paper Award for the Journal on Wireless Communications and Networks, the All University of Oulu Award for research, and the 2019 IEEE ComSoc Radio Communications Committee Early Achievement Award. He is an Editor of the IEEE TRANSACTIONS ON COMMUNICATIONS.
\end{IEEEbiographynophoto}

\end{document}